\documentclass[twocolumn, eqsecnum, aps, showpacs, epsfig]{revtex4}
\usepackage{graphics}
\usepackage{dcolumn}
\begin{document}
\title{\bf{\LARGE{A Study Of  A New Class Of  Discrete Nonlinear
Schr${\ddot{\rm{\bf{o}}}}$dinger Equations}}} 
\author{K. Kundu} 
\affiliation{Institute of Physics, Bhubaneswar 751005, India.}
\date{\today}
\begin{abstract} 
A new class of 1D discrete nonlinear
Schr${\ddot{\rm{o}}}$dinger Hamiltonians with
tunable nonlinerities is introduced, which includes
the integrable Ablowitz-Ladik system as a limit. A new subset of 
equations, which are derived from these Hamiltonians using a generalized
definition of Poisson brackets, and collectively refered to as the N-AL
equation, is studied. The
symmetry properties of the equation are discussed. These equations are
shown to possess propagating localized solutions, having the continuous
translational symmetry of the one-soliton solution of the Ablowitz-Ladik
nonlinear Schr${\ddot{\rm{o}}}$dinger equation.  
The N-AL systems are shown to be suitable to study the combined effect 
of the dynamical imbalance of nonlinearity
and dispersion and the Peierls-Nabarro  potential, arising from
the lattice discreteness, on the propagating solitary wave like profiles.
A perturbative analysis shows that the N-AL systems can have discrete
breather solutions, due to the presence of saddle center bifurcations 
in phase portraits. The unstaggered localized states are shown to have
positive effective mass. On the other hand, large width but small
amplitude staggered localized states have negative effective mass.  
The collison dynamics of two colliding solitary wave profiles are studied
numerically. Notwithstanding colliding solitary wave profiles are seen to
exhibit nontrivial nonsolitonic interactions, certain universal features
are observed in the collison dynamics. Future scopes of this work and 
possible applications of the N-AL systems are discussed.
 
\end{abstract} 
\maketitle

\section{introduction}
As is well known, different nonlinear models can possess spatially
localized solutions for solitary waves\cite{1,2,3}. In many cases, the
solitary waves
are analyzed in the frame work of integrable models which, however,
describe realistic physical systems with certain
approximations\cite{4}.  Two
important examples of integrable nonlinear  equations
are the nonlinear Schr$\ddot{\rm{o}}$dinger (NLS) equation, and
the sine-Gordon
(s-G) equation. While the first one is known to form "dynamical
solitons", the last one yields "kink" and "antikink" solutions. These are
also called
"topological solitons"\cite{1}. Dynamical solitons arise from the acute
balance of
nonlinearity and dispersion. The origin of topological solitons is
the balance of nonlinearity and constraints from topological
invariants\cite{1, 5}. Along side these continuous equations,
a pioneering example of integrable discrete differential equation is
the Ablowitz-Ladik equation. This is often referred to as the
integrable discretization of continuous NLS (ALDNLSE)\cite{6, 7, 8}. On
the other hand, the standard discretization of NLS gives the nonintegrable
discrete nonlinear Schr$\ddot{\rm{o}}$dinger equation (DNLSE)\cite{9,
10, 11}.

The application of these continuous integrable equations in physics is
quite
extensive.  The NLS, which has both dark and
bright solitons, was first used to understand self-focussing and
self-defocussing of carrier waves\cite{12}. This equation later made
inroads in the field of biological energy transport in the guise of
"Davydov's soliton"\cite{2, 13, 14, 15}.  Quite recently, the NLS is
investigated in relation to other phenomena, like Bose-Einstein
condensates in standing waves, discrete solitons and breathers with
dilute Bose-Einstein condensates, and the formation of unstable solitonic
complexes\cite{16, 17}. The
mathematically very rich  s-G equation has also many
applications\cite{1, 2}.  It is used as a model for the propagation
of dislocation in crystals and also for the propagation of
magnetic flux in a long Josephson junction transmission line\cite{1, 2}.
On the other hand, the utility of the ALDNLSE in the analysis of physical
problems is rather tortuous.  Consider for example,
physically motivated models, such as
coupled nonlinear atomic strings with onsite or intersite anharmonic  
potentials\cite{18}, array of coupled optical waveguides\cite{19}, proton
dynamics in hydrogen-bonded chains\cite{4, 14, 20, 21}, the Davydov and
Holstein
models(DHM) for transport in biophysical systems\cite{2, 4, 21}, and so
on. In these models, either the
ALDNLSE does not appear at all or it does not appear in its pristine form.
However, to study the soliton dynamics perturbatively in these models,
the ALDNLSE is the appropriate choice for the zeroth order
approximation\cite{18, 19, 21, 22}. This clearly
shows that the ALDNLSE is an equation of  great deal
of significance in nonlinear science. For other applications
of this equation we mention the following examples. The dynamics of low
frequency and high frequency intrinsic localized modes in nonlinear
lattices can be described to a good approximation by the
ALDNLSE\cite{23, 24}. 
Again, in the study of
dark and bright excitons in systems with exchange and dipole-dipole
interactions, it is shown that in some limiting cases the evolution
equations resulting from model Hamiltonians are reduced to the exactly
integrable ALDNLSE\cite{25}. As another example, we cite the
one-dimensional
Fr$\ddot{\rm{o}}$lich model of exciton(vibron)-phonon interaction. This
model can be approximated by the ALDNLSE, provided we are interested in
the dynamics of large width and small amplitude dynamical
solitons\cite{21}. One other important use of the ALDNLSE lies in the numerical
integration of the NLS. This is done to avoid any numerical instability
problem\cite{6}.

Any small perturbation may break the
integrability so strongly that solitary waves can be unstable and in the
extreme case may altogether disappear. The nonintegrablity in an
otherwise integrable nonlinear equation can also 
give birth to internal
modes, often called "shape modes" in solitary waves. These internal
modes can drastically modify the soliton dynamics\cite{26}. The physical
origin of
integrability breaking terms can be many. For example, this type
of terms can arise from taking into consideration the effect of
thermal fluctuations and the possible absence of order in the
system\cite{8, 27, 28, 29}. However, the most important
one is the discreteness of the underlying space. Consider for example, the
Frenkel-Kontorova (FK) model. It is the spatially discretizd s-G
equation. While the s-G equation which is the long wavelength
approximation of the FK model is integrable, the FK model in its generic
form
is nonintegrable. This nonintegrablity arises from the discretization
reflecting the discreteness of the space\cite{30}. The other interesting 
as well as important example, of course is the already mentioned 
DNLSE\cite{9, 10, 11}. So, relevant in this context
is the study of an IN-DNLS, and also the modified Salerno equation
(MSE)\cite{21,31}. 
The IN-DNLS is a hybrid form of the ALDNLSE and   
the DNLSE with  a "tunable" nonlinearity. On the other hand, in the MSE, 
the usual DNLSE is replaced by a modified version of the DNLSE, the ADNLSE
which involves acousitc phonons, instead of optical phonons in condensed
matter physics parlance\cite{22}. These equations are  studied with
the prime objective to understand the role of lattice discreteness as a
mechanism for the collapse of moving self-localized states to stable, but
pinned localized states\cite{21, 31}. We further note that an IN-DNLS,
which can be continually switched from the DNLSE to the ALDNLSE by varying
a single parameter, is also
studied to investigate the discreteness induced oscillatory instabilities
of dark solitons\cite{32}.

There is at least one continuous nonlinear equation, "$\phi^{4}$
equation", which is nonintegrable, contrary to its "cousin" s-G
system. This $\phi^{4}$ equation can have
either solitary wave solutions or "kink-like" solutions with permanent
profile. However, these solutions do not have the simple collision
properties of solitons. These solutions can bump, lock or annihilate each
other. In addition these always emit some oscillatory disturbances or
radiation in the course of a collision. Furthermore, the head-on
collisions of kink and antikink pair of solutions of
$\phi^{4}$ equation will settle either to a bound state (bion) or to a
two-soliton solution. This settling down is found to depend
fractally on the impact velocity. We further note that all solutions
$\phi^{4}$ equation have profound physical as well as theoretical
significance\cite{33, 34, 35, 36, 37}. In
continuation, we mention that the appearance of fractal
structure in the kink-breather interaction is investigated using the FK
model\cite{30}. Similarly, a fractal structure in solitary-wave collisions
for 
the coupled NLS equations is reported. This structure is observed in the
separation velocity versus collision velocity graph\cite{38}.

On the contary, to the best of my knowledge
there is no known analogy of $\phi^{4}$ type of equation in the discrete
case. Here I propose an extended nonintegrable
version of the ALDNLSE, which has a
"tunable" nonlinearity in the intersite hopping term. At the same time,
the form of nonlinearity is such
that it can allow solitary wave like solutions, as seen in $\phi^{4}$
equation, and its suitable parallel in the linear
regime is one-dimensional correlated disordered
systems\cite{39, 40}. Inasmuch as this nonlinearity is in the intersite
hopping term, it serves two important purposes. First of all, this extra
dispersive correction to the ALDNLSE will try to destroy
the Ablowitz-Ladik (A-L) soliton by dispersion. So, by varying this term
we can investigate the effect of
dispersive imbalance on  the maintance of the moving solitonic
profile. It is relevant at this point to note that both the IN-DNLS and
the MSE
investigate the competition between the on-site trapping and the
solitonic motion of the A-L  solitons. In case of the MSE,
it is found that the narrow A-L solitons, having width smaller than the
critical width will get pinned by the lattice potential\cite{21,
22}. Similar results
are obtained numerically from the IN-DNLS\cite{31}.  Secondly, since the
extra
dispersive term in the proposed equation breaks the integrability of the 
ALDNLSE, the dynamics of the A-L solitons will not be transparent to the 
lattice discreteness. So, along with the IN-DNLS and the MSE, this
model also gives an opportunity to study further the effect of
Peierls-Nabarro (PN) potential on the dynamics of solitons\cite{31,
41}. We
further note
that the DHM yields a very complicated nonlinear discrete differential
equation\cite{21}. So, to gain a better understanding of the DHM, a 
somewhat simpler but physically relevant system needs to be
considered\cite{21}. This is
another important motivation of this study. It is in fact noteworthy in
this context that a slightly modified ALDNLSE has been studied as a
plausible model for dynamical self-trapping in discrete
lattices\cite{42}. As for further motivation of this study, we note that
one rapidly emerging field in nonlinear dynamics is
the study of the solitary-wave interaction in nonintegrable nonlinear
models. The emergence of this field is due to the possibility of observing
many of the predicted effects experimentally, including the soliton energy
and momentum exchange\cite{30, 33, 34, 35, 36, 37, 38, 43, 44}. The model,
that is to be proposed here allows with
an opportunity to study the head-on collision of scalar lattice solitary
waves. Finally, nonlinear phenomena are quite a common occurence in all
branches of natural science. But, our theoretical understanding of
nonlinear equations, particularly that of nonlinear discrete differential
equations is quite limited. So, when viewed from this angle, another
important dimension gets added to the present study.  

The organisation of the paper is as follows. In what follows, I first  
propose a set of discrete nonintegrable Hamiltonians, and derive the
equation of interest from the requisite Hamiltonian.  I study then
analytically the existence of moving solitary
wave like solution in this equation. Since this equation is a
perturbated Ablowitz-Ladik equation, I study this problem further using a
standard perturbative method to find trapped and moving solitons. I
consider next the formation of stationary localized states and the
effective masses of these states. Subsequently, I present some numerical
results to substantiate the analysis. I also study the collison of two
solitary wave profiles numerically using the equation.  Findings
from my study are summarized at the end . In this section I also
discuss the applicabilty and future scopes of the present work.

\section{A general derivation of the Ablowitz-Ladik class of nonlinear 
equations}

We consider here the following Hamiltonian, H.
\begin{widetext}
\begin{eqnarray}
{\rm {H}}\ &=&\ J\ \sum_{n}\ (\phi^{\star}_{n}\ \phi_{n+1}\ +\
\phi^{\star}_{n+1}\ \phi_{n})\ -\ \frac{1} {2}\ \sum_{n}\ \sum_{j\ =\
1}^{l}
g^{j}_{0}\ (\phi^{\star}_{n}\ \phi_{n+j}\ +\
\phi^{\star}_{n+j}\ \phi_{n})^{2}\ + \frac{2 \nu} {\lambda}\ \sum_{n}
|\phi_{n}|^{2}\nonumber\\
 &-&\ \frac{2} {\lambda}\ (\frac{\nu} {\lambda}\ -\
J)\ \sum_{n} \ln{[1\ +\ \lambda\ |\phi_{n}|^{2}]}.     
\end{eqnarray}
\end{widetext}
\noindent
It is understood that the Hamiltonian, ${\rm {H}}$ in Eq.(2.1) describes a
conservative system in which a quasiparticle (exciton, vibron etc.) moves
in an one-dimensional chain.  To obtain the time evolution equations
for generalized coordinates,
$\{\phi_{m},\ \phi^{\star}_{m} \}$, we make use of the nonstandard
Poisson bracket relationship among these generalized coordinates as
given in Appendix A. After defining
$\phi_{m}\ =\ (-1)^{m}\ \psi_{m}$, we get for $m\ =\ 1,2,\
\dots $
\begin{widetext}
\begin{eqnarray}
{\mathcal{F}}_{j} (\psi_{m}, g^{j}_{0})\ &=&\ g^{j}_{0}\
\sum_{\sigma\ =\ \pm 1} (\psi^{\star}_{m}\
\psi_{m + \sigma j}\ +\ \psi^{\star}_{m + \sigma j}\ \psi_{m})\ \psi_{m
+ \sigma j}\nonumber\\
{\mathcal {F}}(\psi_{m}, \lambda)\ \ \ &=&\ \ \ (1\ +\ \lambda\
|\psi_{m}|^{2})\
\sum^{l}_{j\ =\ 1} {\mathcal{F}}_{j} (\psi_{m}, g^{j}_{0}) \\
i\  \dot{\psi}_{m}\ -\ \ & 2 J & \ \psi_{m}\ +\ J\ (1\ +\
\lambda\ |\psi_{m}|^{2})\ (\psi_{m+1}\ +\ \psi_{m-1})\
 =\ \  2\ \nu\ |\psi_{m}|^{2}\ \psi_{m}\  -\  {\mathcal {F}}(\psi_{m},
\lambda).
\end{eqnarray}
\end{widetext}
In ${\mathcal{F}}(\psi_{m}, \lambda)$, arguements, $g^{0}_{j},\  j\ =\
1, 2, \cdots,\ l $ have been suppressed, and $J\ >\ 0$.  When $l \ =\ 1$,
in Eq.(2.1) and
consequently in Eq.(2.3) we have
just an extra nonlinear nearest-neighbor coupling in hopping with a
coupling
constant, $g^{1}_{0}$. This coupling is assumed to arise due to
quasiparticle-phonon
interaction\cite{21}. Similarly, when $l\ =\ 2$, we have both nonlinear
nearest-neighbor and next-nearest-neighbor couplings. Coupling constants
are $g^{1}_{0}$ and  $g^{2}_{0}$ respectively. So, 
in this model coupling constants are assumed to depend on the
distance between the sites involved. For any arbitrary $l$ then, 
a given site, over and above the standard linear nearest-neighbor
coupling, is
coupled through nonlinear hopping to $l$ consecutive sites in general on
both sides
of it. Of course, the origin of such coupling is assumed to arise due to
the heuristic generalization of quasiparticle-phonon interaction beyond
the nearest neighbor. We further note that in this model also
\begin{equation}
{\mathcal{N}} =\ \sum^{\infty}_{ m =\ -\infty} \ln {[1 +\ \lambda\
\ |\psi_{m}|^{2}]}\ \ =\ {\rm {Constant}}
\end{equation}     
as in the original Ablowitz -Ladik equation (see also appendix A). So, if
$\lambda\
|\psi_{m}|^{2}\
\ll\ 1$ for all m, we get $\sum_{m}|\psi_{m}|^{2}\ \approx {\rm
{Constant}}$.   
We now consider the following limits of Eq.(2.3).
(1) When $g^{j}_{0}\ =\ 0$ for all $j$, we have the Salerno Equation in
the quantum version. In the classical domain, it has been nomenclatured as
IN-DNLS and the formation of staggered localized states from this equation
is studied\cite{31, 45, 46}.
(2) When $\nu\ =\ 0$ as well $g^{j}_{0}\ =\ 0 $ for all $j$, we have the
standard Ablowitz-Ladik equation. This equation is known to have a single
soliton solution\cite{6, 7}.
(3) When $\lambda\ =\ 0\ =\ \nu$ and also $l\ =\ 1$, we have the model
where exciton-phonon interaction affects only the hopping between
nearest-neighbors. Of course, to obtain this particular nonlinear
mathematical form of
interaction, it is assumed that the lattice relaxation is faster 
than the quasiparticle
dynamics\cite{21}. This is a very  standard assumption which is used in
the
Davydov's soliton\cite{13, 22} and 
also in the Rashba and Toyozawa mechanism for the formation of
self-trapped exciton\cite{47, 48, 49}. This equation is also studied for
single soliton
solution, using perturbation method\cite{21}. 
(4) The equation with $\nu\ =\ 0$ will be refered to here as the
nonintegrable Ablowitz-Ladik (N-AL) equation. Consider again the situation
where we take $l\ =\ 1$ in the N-AL equation and further ignore altogether
in Eq.(2.3) the term having
$g^{1}_{0}\ \lambda$. We have then a model equation which
describes a truncated version of the model, in which
exciton(vibron)-phonon interaction affects both site-energy and hopping. 
We note that a perturbative analysis of the full model
is done for one soliton solution\cite{21}. Here our plan is to analyze 
various aspects of the N-AL equation. 

Before we proceed furthere, we define
$\tau\ =\ Jt,\ \ \Psi_{m}\ =\ \sqrt{\lambda}\  \psi_{m},\  
m\ \in Z,\ {\rm {and}}\ g_{j}\ =\ \frac{g^{j}_{0}} {\lambda\ J}.$ We
further note that ${\mathcal{F}}_{j}(\frac{\Psi_{m}}
{\sqrt{\lambda}}, g^{j}_{0})\ =\ \frac{1} {\sqrt{\lambda}}\
{\mathcal{F}}_{j}(\Psi_{m}, g_{j}),\ \ j\ =\ 1,2,\ \cdots\ l $. This, in
turn yields ${\mathcal{F}}(\frac{\Psi_{m}}
{\sqrt{\lambda}},\ \lambda)\ =\ \frac{1} {\sqrt{\lambda}}\ {\mathcal{F}}
(\Psi_{m}, 1),\ m\ \in Z$. So, in the limit of $\nu\ =\ 0$,
we have from Eq.(2.3) and $m \in Z$
\begin{eqnarray}
i\  \dot{\Psi}_{m}\ -\  2 \ \Psi_{m}\ +\  (1\ &+&\ |\Psi_{m}|^{2})
\ (\Psi_{m+1}\ +\ \Psi_{m-1})\nonumber\\
&=&\ -\  {\mathcal {F}}(\Psi_{m}, 1)
\end{eqnarray} 
and Eq.(2.4) also remains valid with $\lambda\ =\ 1$ and $\psi_{m}$
replaced by $\Psi_{m}$. Again, ${\mathcal{F}}(\Psi_{m},1)$ should have in
its arguement $g_{1},\ g_{2}, \cdots,\ g_{l}$, which have been suppressed
for convenience. We note that Eq.(2.5) possesses a reflection symmetry. To
understand this, we consider the transformation, $\Psi_{m}\ \rightarrow\
(-1)^{m}\ \Psi_{m}\ \exp[- 4 i \tau ]$. The equation will remain invariant
under this transformation provided  $\tau\ \rightarrow\ - \tau $ and
$g_{j}\ \rightarrow\ - g_{j},\ \ j\ \in \ l$.  Again, if
${\mathcal{F}}(\Psi_{m}, 1)\ =\ 0,\ m\
\in\ Z$, the equation becomes self-dual under this reflection
transformation.

\section{Propagating solutions of Eq.(2.5)}

In order to gain an understanding of the nature of the solution of
Eq.(2.5), we consider the case where  $g_{j}\ =\ 0$ for all $j$.  As
mentioned earlier, in
this limit we have the integrable Ablowitz-Ladik discrete nonlinear
Schr$\ddot{\rm o}$dinger equation(ALDNLSE)\cite{6, 7}. The exact
one-soliton solution
of the ALDNLSE, $\Psi^{0}_{m}$, $m\ \in Z$ is
\begin{eqnarray}
\Psi^{0}_{m}(\tau)\ &=&\ \frac{\sinh{\mu}} {\cosh[\mu (m-x)]}
\exp[ik(m-x)-i\alpha]\nonumber\\
&=&\ Q_{m}(\tau)\ \exp[ik(m-x)-i\alpha]
\end{eqnarray}
with the following equations for the soliton parameters:
\begin{eqnarray}
\dot{\mu} &=&0,\ \dot{k}=0,\ \dot{\alpha} = \omega,\\
\omega\ &=&\ 2\ [1-\cosh{\mu} \cos{k}],\\
\dot{x}\ & =&\ \frac{2} {\mu} \sinh{\mu}\ \sin{k}.
\end{eqnarray}
So, for each $\mu$ there exists a band of velocities [see Eq.(3.4)] at
which the localized state or the one-soliton state can travel without
experiencing any PN pinning due to the lattice
discreteness\cite{31, 41}. We also note that for a given $k$, $\mu\ \in
[0, \infty]$.
Mathematically, one soliton solution of the ALDNLSE describes two
parameters,
namely $k\ {\rm {and}}\ \mu$, family of curves. So, even if we pin the
value of one of those parameters to a prescribed value, Eq.(3.1) will
still be a solution of the ALDNLSE. 

We now consider ${\mathcal{F}}(\Psi^{0}_{m}, 1)$. Introducing Eq.(3.1) in
Eq.(2.5) we get
\begin{widetext}
\begin{equation}
\frac{{\mathcal{F}}(\Psi^{0}_{m}, 1)} {2\ (1 + Q_{m}^{2})\
\Psi^{0}_{m}(\tau)}\ =\  \sum^{l}_{j\ =\ 1}
\{g_{j}\ \cos{kj}\  [( Q_{m+j}^{2}\ +\ Q_{m-j}^{2})\ \cos{k j}\
\ +\ i(Q_{m+j}^{2}\ -\
Q_{m-j}^{2})\ \sin{k j}]\}.
\end{equation}
\end{widetext}
For convenience in further discussion, we consider only one term, say the
l-th term of the sum in Eq.(3.5). We note that for $\cos{k l}\ =\ 0$,
permissible values of $k$ are $k\ =\ \pi\ -\ (\frac{2 j_{1} + 1} {l})\
(\frac{\pi} {2}),\ \ j_{1}\ =\ 0, 1, 2, \cdots, (2 l -1)$. So, there are
$2 l$ permissible values of $k$ for $k\ \in [- \pi, \pi]$. We now state 
the following results.

(1) For $l\ =\ 1$, ${\mathcal{F}}(\Psi^{0}_{m}, 1)\ =\ 0\ {\rm{for}}\ m\
\in Z$, if $|k|\ =\ \frac{\pi} {2}$. So, for this case Eq.(3.1) with
$k\ =\ \pm \frac{\pi} {2}$  are the solitary wave like solutions of
Eq.(2.5). We further note that solitary waves have the maximum possible
speed.

(2) $l$ is odd as well only odd values of $j$ in the sum in the
Hamiltonian, ${\rm {H}}$ are permissible. Also,in this case, only $|k|\ =\
\frac{\pi} {2}$ is allowed. So, Eq.(3.1) with these particular
values of $k$ are the solitary wave like solutions of Eq.(2.5).

(3) $l$ is even as well only even values of $j$ in the sum in Eq.(3.5) are
permissible. In this case, there is no permissible value of $k$. So,
Eq.(2.5) will not have any solitary wave like solution.

(4) $l$ is arbitrary and $j$ takes odd values with at least one even
value. In this case, even if $l$ is odd, there is no permissible value of
$k$. So, Eq.(2.5) has no solitary wave like solution.

(5) There is only one term, say $l$-th term in the sum in Eq.(3.5). In
this
situation, Eq.(3.1) describes the solitary wave like solution of
Eq.(2.5) with $2 l$ permissible values of $k$ which are already given.

We also note that when $|g_{j}|\ <<\ 1 $ for $j\ =\ 1, 2, \cdots\ l$,
Eq.(2.5) then can be treated as a
perturbed ALDNLSE. In this situation we can use the standard perturbation
theory to investigate the soliton dynamics from Eq.(2.5). This aspect is
considered next.

\section{The perturbative soliton dynamics of Eq.(2.5)}
The method is amply discussed in the literature\cite{21, 22, 50}. In this
method, it is
assumed that the zeroth order solution is the standard one-soliton
solution of the ALDNLSE. It is further assumed that the effect of
perturbation
on the soliton dynamics can be adequately taken into consideration by
allowing four parameters, namely $x$, $k$, $\mu$ and $\alpha$ to vary
adiabatically in time, $\tau$.  Application
of the method to the N-AL equation gives for $\mu$, $k$ and $x$ :
\begin{eqnarray}
S(\mu, x)\ &=&\ \sum^{\infty}_{s\ =\ 1} \frac{\frac{\pi^{2} s} {\mu}}
{\sinh{\frac{\pi^{2} s} {\mu}}}\ \cos{2 \pi s x}\nonumber\\
G(k, l, \mu)\ &=&\ \sum^{l}_{j\ =\ 1} g_{j}\ \frac{\ {\cos^{2}}{k j}}
{{\sinh^{2}}{\mu j}}\nonumber\\
G_{1}(k, l, \mu)\ &=&  4\ \frac{{\sinh^{4}}{\mu}} {\mu^{2}}\ [\mu\ \coth{\mu}
\ -\ 1 - 2\ S(\mu, x)]\nonumber\\ 
G_{2}(k, l, \mu)\ &=&\ G_{1}(k, l. \mu)\  \frac{\partial G(l, k, \mu)}
{\partial k\ \ \ \ \ \ \ \ \ \ }\nonumber\\  
\dot{\mu}\ \ \ \ = \ &0& \\
\dot{k}\ =\ -\ &8&\  \frac{{\sinh^{4}}{\mu}} {\mu^{2}}\ G(k, l, \mu)\
\frac{\partial S(\mu, x)} {\partial x\ \ \ \ \ \ \ \ }\\
\dot{x}\ \  =\ - &2&\ \frac{\sinh{\mu}} {\mu}\ \frac{\partial \cos{k}}
{\partial k\ \ \ \ \ \ }\ -\ G_{2}(k, l, \mu).
\end{eqnarray} 
We first note thar the famous Poisson sum formula is used to obtain
$S(\mu, x)\ {\rm{and}}\ G(k, l, \mu)$\cite{21}. 
In the situation where $\cos{k j}\ =\ 0$ for all $j$, both $G(k, l, \mu)$
and $\frac{\partial G(l, k, \mu)} {\partial k\ \ \ \ \ \ \  }$ are
zero. In this case, it is easy to see that  $\dot{\mu}$, $\dot{k}$ and
$\dot{x}$ are given by Eq.(3.2) and Eq.(3.4) respectively. Of course, for
$\dot{x}$ only certain values of $k$ for $k\ \in [-\pi, \pi]$ are allowed.
Notwithstanding this, this case is as expectedly similar to the
one-soliton solution of the ALDNLSE.  We consider next the problem of 
an effective Hamiltonian. 

\subsection{An effective Hamiltonian}
In order to derive an effective Hamiltonian for the
dynamical
system described by Eqs.(4.2) and (4.3), we mutiply these two equations by
$\dot{x}$ and $\dot{k}$ respectively. Now, substracting the first one from
the second one, we get
\begin{widetext}
\begin{equation}
- [2\ \frac{\sinh{\mu}} {\mu}\ \frac{\partial \cos{k}}
{\partial k\ \ \ \ \ \ }\ + \
4\ \frac{{\sinh^{4}}{\mu}} {\mu^{2}}\ [\mu\ \coth{\mu}\ -\ 1 - 2\ S(\mu,
x)]\ \frac{\partial G(l, k, \mu)} {\partial k\ \ \ \ \ \ \ \ \ \
}]\ \dot{k}\ 
+\ 8\  \frac{{\sinh^{4}}{\mu}} {\mu^{2}}\ G(k, l, \mu)\
\frac{\partial S(\mu, x)} {\partial x\ \ \ \ \ \ \ \ }\ \dot{x}\ =\ 0.
\end{equation}
\end{widetext}
It is easy to see that Eq.(4.4) defines a constant of motion of the
dynamical system. This constant can be called the effective Hamiltonian,
$H_{\rm {eff}}(x, k, \mu, l)$ of the system and from Eq.(4.4) we get
\begin{widetext}
\begin{equation}
H_{\rm {eff}}(x, k, \mu, l)\ =\ - 2\ \frac{\sinh{\mu}} {\mu}\
\cos{k}
\ -  \
4\ \frac{{\sinh^{4}}{\mu}} {\mu^{2}}\ [\mu\ \coth{\mu}\ -\ 1 - 2\ S(\mu,
x)]\  G(l, k, \mu).
\end{equation}
\end{widetext}
Again, it is easy to see from Eq.(4.4) that
\begin{equation}
\dot{x}\ =\ \frac{\partial H_{\rm{eff}}} {\partial k\ \ \ \ }\ \ {\rm
{and}}\ \  \ \dot{k}\ =\ -\ \frac{\partial H_{\rm{eff}}} {\partial x\ \
\ \ }
\end{equation}
together yield the original equations of $\dot{x}$ and $\dot{k}$. In the 
subsequent analysis, the $l\ =\ 1$ case being the physically most relevant
one, is analyzed. In another
simplification, $S(\mu, x)$ is approximated by its the most dominant
first term. This simplification will not bring any qualitative change in
the dynamics.

\subsection{The analysis of fixed point} 
To obtain the fixed points of the set of equations, namely  Eq.(4.2) and
Eq.(4.3), we set $\dot{x_{s}}\ =\ 0\ =\ \dot{k_{s}}$. This then gives
$(x_{s}, k_{s})\ =\ (\frac{p} {2}, n \pi)$ where $ p\ {\rm {and}}\ n\ =\
0,\  \pm 1,\ \pm2,\  \cdots$. To find the phase portraits around these
fixed
points, we define $ z_{1}\ =\ x\ -\ x_{s}$ and $ z_{2}\ =\ k\ -\ k_{s}$. 
We define for convenience and clarity
\begin{eqnarray}
A_{0}(\mu)\ &=& \ \frac{\sinh{\mu}} {\mu}\nonumber \\
A_{1}(\mu)\ &=& \ A^{2}_{0}(\mu)\ (\mu\ \coth{\mu}\ -1)\nonumber \\
A_{2}(\mu)\ &=& \ A^{2}_{0}(\mu)\ \frac{\frac{\pi^{2}} {\mu}}
{\sinh{\frac{\pi^{2}} {\mu}}}.
\end{eqnarray} 
We further define
\begin{eqnarray}
B_{0}(\mu, p)\ &=&\  A_{1}(\mu)\ -\ (-1)^{p}\ 2\  A_{2}(\mu)\nonumber \\
B_{1}(\mu, g_{1}, p, n)\ &=& \  (-1)^{n} A_{0}(\mu)\ +\ 4\  g_{1}\ 
B_{0}(\mu,p)\nonumber\\ 
B_{2}(\mu, g_{1})\ & =& \ 16\  \pi^{2}\ g_{1}\  A_{2}(\mu).
\end{eqnarray}

Now, because of our stated assumptions, we get
\begin{eqnarray}
\dot{z_{1}}\ &=& \ 2\  B_{1}(\mu, g_{1}, p, n)\  z_{2}\nonumber\\
\dot{z_{2}}\ & =& \ (-1)^{p}\  2\  B_{2}(\mu, g_{1})\ z_{1}.
\end{eqnarray}
This set of equations, in turn yields
\begin{equation}
\frac{z_{1}^{2}} {B_{1}(\mu, g_{1}, p, n)}\ -\ (-1)^{p}\ \frac{z_{2}^{2}}
{B_{2}(\mu, g_{1})}\ =\ {\rm {Constant}}.
\end{equation}
From Eq.(4.10), it is transparent that the system has only two types of
fixed points. These are centres and saddle hyperbolic fixed
points. We give below some in depth analysis of fixed points.

We note that $A_{i}(\mu),\ i\ =\ 0, 1, 2$ are even fuctions of $\mu$.
This, in turn implies that $B_{i},\ i\ =\ 0, 1, 2$ are also even functions
of $\mu$. We take $g_{1}\ >\ 0$ in this analysis. There will be no loss of
generality by this assumption. We now note the followings.\\
(i) $B_{0}(\mu, p)\ \rightarrow\ 0,\ {\rm{as}}  \ \mu\ \rightarrow\
0$. It
can also be shown that $B_{0}(\mu, p)$ is a monotonically increasing
function of $\mu$. In other words, it is a monotonically increasing
positive semidefinite function of $\mu$. These properties are true,
irrespective of p being even or odd.\\
(ii) $B_{1}(\mu, g_{1}, p, n)\ \rightarrow\ (- 1)^{n},\ {\rm {as}}\ \mu\
\rightarrow\ 0$. Furthermore, it can be seen by plotting this function
against $\mu$ that $B_{1}(\mu, g_{1}, p, n)\ > 0 $, when $\mu\
\rightarrow\
\infty$. Both properties are true irrespective of the nature of
p. Furthermore, when n is even, $B_{1}(\mu, g_{1}, p, n)$ is a
monotonically
increasing as well as a positive definite function of $\mu$. On the other
hand, when n is odd, there exists a value of $\mu$, say $\mu_{r}(g_{1},
p)$ such that $B_{1} (\mu_{r}, g_{1}, p, n)\ =\ 0$. Again, $\mu_{r}(g_{1},
p)$
has the following properties. (a) For a given value of
$g_{1}$,\ $\mu_{r}(g_{1},\ even p)\ >\ \mu_{r}(g_{1},\  odd
p)$, (b) irrespective of the nature of p, the value of $\mu_{r}$ 
monotonically decreases with increasing $g_{1}$, and (c) $\mu_{r}(g_{1},\ 
even p)\ \rightarrow\ \mu_{r}(g_{1},\  odd p)\ {\rm {as}}\  g_{1}\
\rightarrow\ \infty$. See also Fig.1.\\
\begin{figure}
\parindent 0.3in
\includegraphics{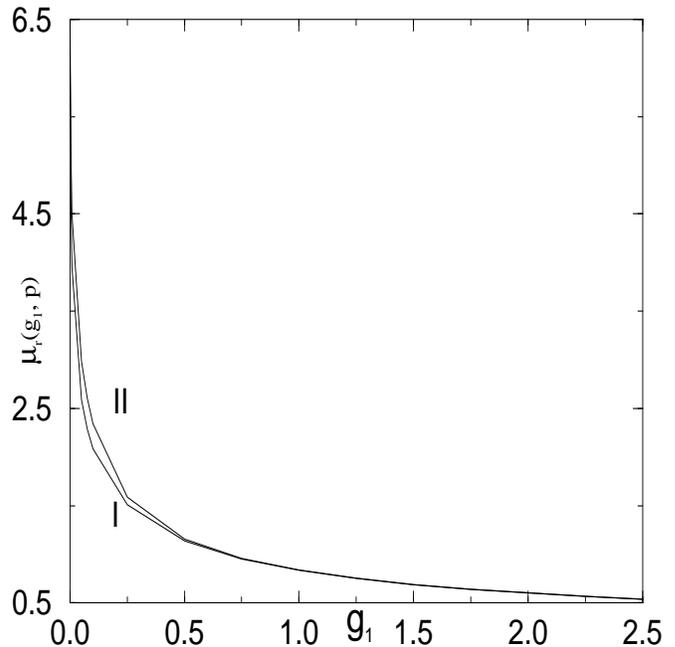}
\caption{This shows the variation of $\mu_{r}(g_{1}, p)$ as a
function of
$g_{1}$ where $\mu_{r}$ is the root of $B_{1}(\mu, g_{1}, p,
n)$. Eq.(4.8) in the text. $n$ is odd. I: p is odd and II: p is even.}
\end{figure}
(iii) $B_{2}(\mu, g_{1})$ is a monotonically increasing positive
semidefinite function of $\mu$.\\
We now consider case by case to determine the nature of fixed points from
the linearized analysis. We note that the classifications are done
using standard analysis\cite{51, 52}.\\
Case I : n and p are both even. In this case, both $B_{1}(\mu, g_{1}, p, 
n)$
and $B_{2}(\mu, g_{1})$ are positive. Eq.(4.10) defines a hyperbola. So,
the
fixed point is a saddle point.\\
CaseII ; n is even, but p is odd. Since, the nature of these two functions
remains the same, Eq.(4.10) defines an ellipse. So, we have a 
center.\\
Case III: Both n and p are odd. Again, $\mu\ >\ \mu_{r}(g_{1}, even
p)$. In this case, both $B_{1}$ and $B_{2}$ are positive. So,
Eq.(4.10) defines an ellipse and we have a center.\\
Case IV: n is odd, but p is even.  Furthermore, $\mu\ >\ \mu_{r}(g_{1},
even p)$. In this case also, both $B_{1}$ and $B_{2}$ are positive. But
due to
the evenness of p, Eq.(4.10) defines a hyperbola and we have a saddle
point.\\
Case V: Both n and p are odd, but  $\mu\ <\ \mu_{r}(g_{1}, odd
p)$. In this case, while $B_{2}$ is  positive, $B_{1}$ is negative. So,
Eq.(4.10) defines a hyperbola and the fixed point is a saddle point.\\
Case VI:  n is odd, but p is even. Again,  $\mu\ <\ \mu_{r}(g_{1}, odd
p)$. In this case, Eq.(4.10) defines an ellipse. So, we have a center.\\ 
Case VII: Both n and p are odd. Again,  $\mu_{r}(g_{1}, odd p)\ < \mu\ <\
\mu_{r}(g_{1}, even
p)$. In this case, both $B_{1}$ and $B_{2}$ are positive. Since, p is odd, 
Eq.(4.10) defines an ellipse. So, we have a center.\\ 
Case VIII: n is odd, but p is even.  Furthermore,  $\mu_{r}(g_{1}, odd p)\
< \mu\ <\
\mu_{r}(g_{1}, even
p)$. In this case, while $B_{1}$ is negative, $B_{2}$ is, however,
positive. Since, p is even,  Eq.(4.10) defines an ellipse. So, we have a
center.\\ 
Our analysis is tabulated below. By Region I, it is meant that $\mu \ <\
\mu_{r}(g_{1}, odd p)$. Region II and III are respectively characterized
by $\mu_{r}(g_{1}, odd p) \ <\ \mu \ <\ \mu_{r}(g_{1}, even p)$ and $\mu
\ >\ \mu_{r}(g_{1}, even p)$.
\newpage   
\begin{center}
TABLE I : Nature of fixed points from linearization analysis
\end{center}

\begin{center}
\begin{tabular} {|ccc|ccc|ccccccccccccccc|} \hline
\multicolumn{3}{|c|}{$ x_{s}\ =\ \frac{p} {2}$}
 &\multicolumn{3}{c|} {$k_{s}\ =\ n\ \pi$}    & 
& & &\multicolumn{12}{c|}{
\ \ \ \ \ \ \ Nature of fixed points\ \ \ \ \ \ \ \ } \\
\hline 
\multicolumn{3}{|c|}{p} &\multicolumn{3}{c|} {n}
&\multicolumn{5}{c|}{\ \ Region I\ \ } &\multicolumn{5}{c|}{\ \ Region
II\ \ }  
&\multicolumn{5}{c|}{\ \ Region III\ \ }\\ 
\hline
\multicolumn{3}{|c|}{\ \ \ even\ \ \ } &\multicolumn{3}{c|}{\ \ \ even\ \
\ 
}
&\multicolumn{5}{c|}
{saddle} &\multicolumn{5}{c|}{saddle} &\multicolumn{5}{c|}{saddle}\\
\hline
\multicolumn{3}{|c|}{odd} &\multicolumn{3}{c|}{even} &\multicolumn{5}{c|}
{\ \ \ center\ \ \ } &\multicolumn{5}{c|}{\ \ \ center\ \ \ }
&\multicolumn{5}{c|}{\ \ \ center\ \ \ }\\
\hline
\multicolumn{3}{|c|}{odd} &\multicolumn{3}{c|}{odd} &\multicolumn{5}{c|}
{saddle} &\multicolumn{5}{c|}{center}
&\multicolumn{5}{c|}{center}\\
\hline
\multicolumn{3}{|c|}{even} &\multicolumn{3}{c|}{odd} &\multicolumn{5}{c|}
{center} &\multicolumn{5}{c|}{center}
&\multicolumn{5}{c|}{saddle}\\
\hline
\end{tabular}
\end{center}
\subsection{A numerical analysis of phase portraits}
Regarding the numerical investigation of phase diagrams around fixed
points, we note that there are all together six fundamental
possibilities. Two possibilities arise from $k_{s}$ being even or
odd. Again, for each case, there are three possibilities, depending on the
magnitude of $\mu$. In this discussion, we consider $k_{s}\ =\ 0\ {\rm
{and}}\  \pi$. Along the line, $k_{s}\ =\ 0$, we have a set of fixed
points
at $x_{s}\ =\ 0,\ \pm \frac{1} {2},\ \pm 1,\ \pm \frac{3} {2},\ \pm 2,
\cdots$. According to the analysis, based on linearization $x_{s}\ =\ 0,\
\pm 1,\ \pm 2,\ \cdots $ should be saddle hyperbolic fixed points.
Between every two consecutive saddle points, one should
naturally expect centers. This is perfectly borne out in
the numerical investigation (Fig.2). 
\begin{figure}
\parindent 0.25in
\includegraphics{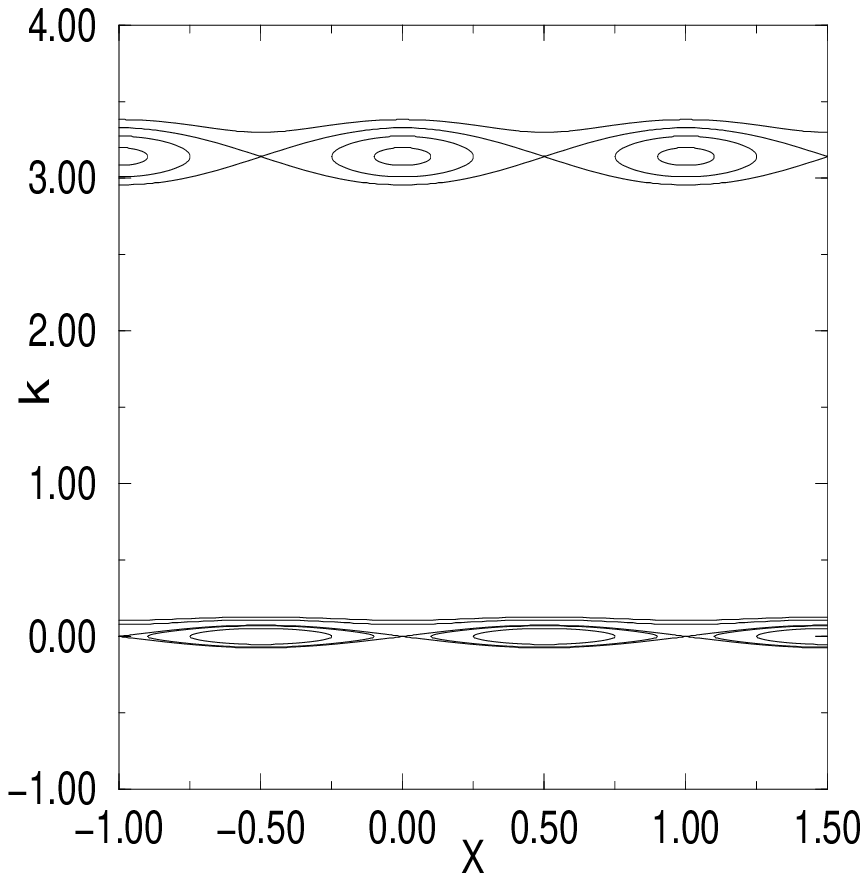}
\caption {The bottom part of the figure is the  phase portrait
that shows
the trapped and the moving solitons in
Region I for the fixed points having  $k_{s} = 0.0$.  For moving
solitons $k_{m0}\ =\ \frac{\pi} {40}\ {\rm{and}}\ \frac{\pi} {30}$
respectively. The upper part of the figure is the  phase portrait
that shows the trapped
and the moving solitons in Region I, but for the fixed points having
$k_{s} = \pi$. For the moving soliton $k_{m \pi}\ =\ \frac{21 \pi} {20}$.
$\{x_{s}\}$ are shown in the figure.  $l\ =\
1, g_{1}\ =\ 0.5,\ {\rm{and}}\ \mu\ =\ 1.0$ for all curves in the figure.
Eq.(4.5) is used to obtain this phase portrait. See also the text.}
\parindent 0.3in
\includegraphics{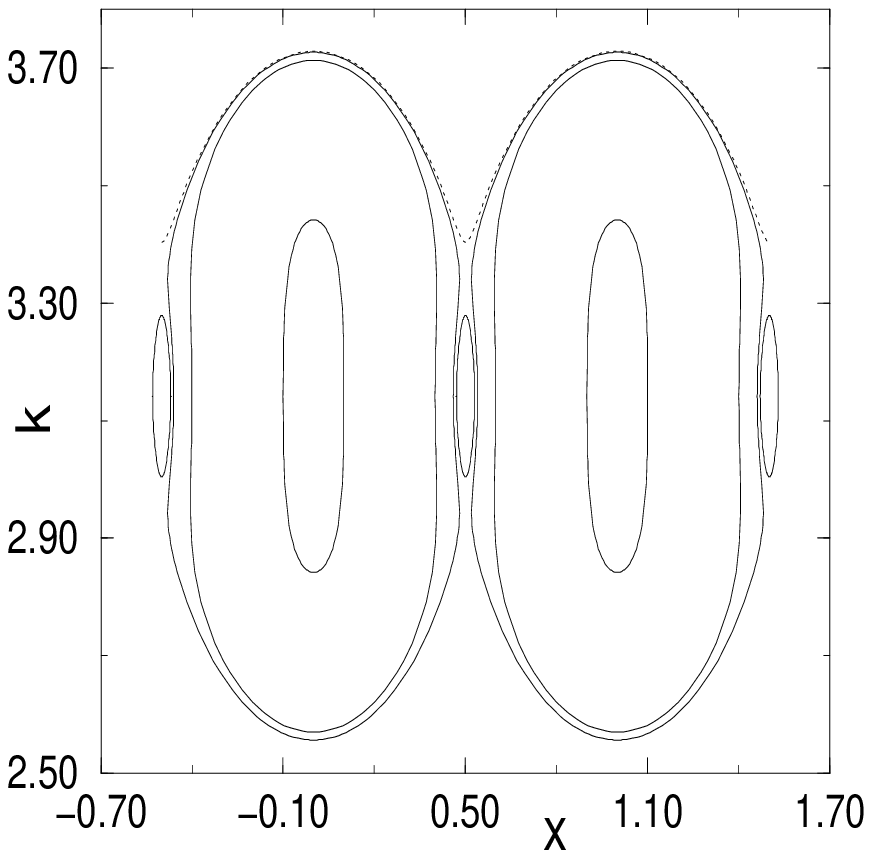}
\caption{ This phase portrait shows the trapped and the moving   
solitons in
Region II for fixed points having $k_{s} = \pi$.
$\{x_{s}\}$ are, of course, shown
in the figure. As per the text
$l\ =\ 1$. Other parameters, namely $ g_{1}\ =\ 0.5,\ {\rm{and}}\ \mu\ =\
1.15$. For the moving soliton $k_{m\pi}\ =\  \frac{13 \pi} {12}$.}
\end{figure}
\begin{figure}
\parindent 0.35in
\includegraphics{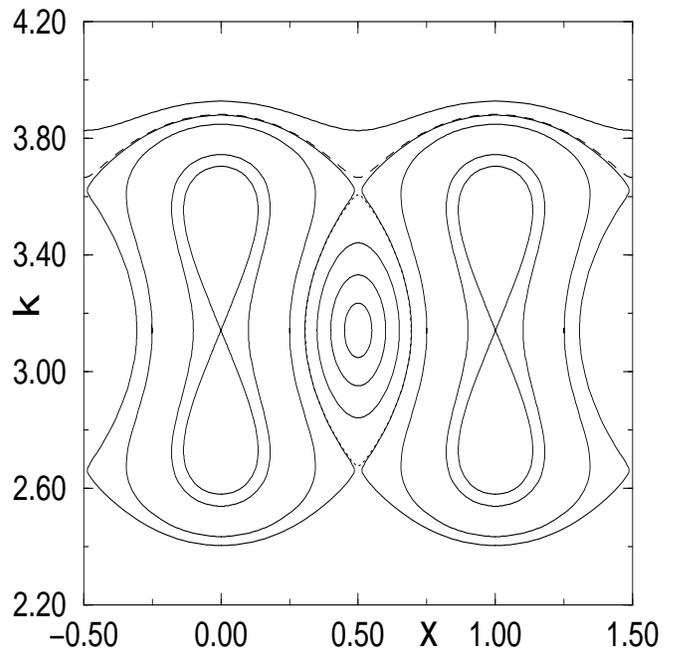}
\caption{ This phase portrait shows the trapped as well as the
moving
solitons in Region III for fixed points having $k_{s} = \pi$.    
$\{x_{s}\}$ are, of course, shown
in the figure. As per the text
$l\ =\ 1$. Furthermore, $ g_{1}\ =\ 0.5,\ {\rm{and}}\ \mu\ =\ 1.2$.
For the moving soliton $k_{m\pi}\ =\  \frac{7 \pi} {6}\ {\rm{and}}\
\frac{5 \pi} {4}$.}
\end{figure}
It is also to be noted that
Fig.2 shows the phase
diagram in the region I ($\mu\ <\ \mu_{r}(g_{1}, oddp)$).  
To understand the origin of 
a saddle fixed point between two consecutive centers, 
we note that two
outermost ellipses encircling two
centers will definitely touch at a point on the
$x$-axis. Then, the flow of phase curves  emanating from this
point as well as in the vicinity of it will indicate a
saddle hyperbolic fixed point. By the similar type of argument, the
appearance of a center between two consecutive saddle fixed
points can be understood.  It is also found that the
absolute critical value of $k$, below which localization of soliton
occurs, increases monotonically with increasing $\mu$. This behavior is
consistent with the physics of the problem. 

We consider now the case of $k_{s}\ =\ \pi$. It has certain interesting
twists, which is quite a common occurence in the dynamics of nonlinear 
systems\cite{52}. In the region I,
that is, $\mu\ <\ \mu_{r}(g_{1}, oddp)$, according to our standard
linearization based analysis, $x_{s}\ =\ 0, \pm 1,\ \pm 2,\cdots$ should
be centers. On the other hand, $x_{s}\ =\ \pm\frac{1}
{2},\ \pm \frac{3} {2},\cdots $ should be saddle hyperbolic fixed points.  
The reason to have a saddle hyperbolic fixed point between two consecutive
centers has already been put forward in the previous paragraph.
Our numerical investigation also confirms this result (Fig.2). In the
region II
($\mu_{r}(g_{1}, oddp)\ <\ \mu\ <\ \mu_{r}(g_{1}, evenp)$), the
linearization
analysis tells us that all fixed points along the line $k_{s}\ =\ \pi$ are
centers. To understand this scenario, we note the following. Consider
two centers, which are next but one to each
other. Outermost ellipses, encircling these fixed points can also
intersect. We then have a convex lens type region in the phase space
(Fig.3).
Within this region of phase space, phase curves will have no other
option but to close around a point on the $x$-axis. This fixed
point will also then be another center. This is clearly
seen in Fig.3, which describes the phase diagram of the soliton dynamics
in the region II. 
 
It is needless to say that the phase diagram is very
interesting and unique. To the best of my knowledge, this is the first
case
where a chain of centers with no hyperbolic fixed point 
is found. The phase diagram in the region III is shown in Fig.4. 

This region is defined by $\mu\ >\ \mu_{r}(g_{1}, evenp)$. According to
the linearization
analysis, along the line $k_{s}\ =\ \pi$, $x_{s}\ = \pm \frac{1} {2},\
\pm \frac{3} {2},\cdots$ should be centers. This is
indeed found in the numerical investigation. Again, the same linearization
analysis tells that $x_{s}\ =\ 0,\ \pm 1,\ \pm 2, \cdots$ will be
saddle hyperbolic fixed points. From the numerical study with $\mu\ =\
1.2\ {\rm{and}}\ g_{1}\ =\ 0.5$, we find that
upto a certain distance from each of these fixed points, phase curves tend
to diverge from the fixed points, indicating their hyperbolic
structure. However, after a critical value of $x$, phase curves turn back
to make close curves. This happens upto a certain maximum value of $x$
from the relevant fixed point on
the line $k_{s}\ =\ \pi$. After that value of $x$, phase curves close
around the nearest center (Fig.4). In case of $x_{s}\ =\
0$, it is found that the phase curve with the intial value of $x$, $x_{0}\
=\ 0.306$ closes around $x_{s}\ =\ 0$. But, for $x_{0}\ =\ 0.307 $, the
phase curve  encircles $x_{s}\ =\ 0.5$. So, for these two fixed points,
the transition point is somewhere in between these two values. 

It is important to realize that in this part of the problem we find three
important phenomena that are encountered in nonlinear dynamics, namely
(i) bifurcation, (ii) discrete breathers and (iii) relaxation
oscillations\cite{51, 52, 53, 54}. We note that
for $\mu\ >\ \mu_{r}(g_{1}, odd p)$, that is, in  regions II and III, all
saddle points are changed to
centers (Fig.3 and Fig.4). So, we obtain a bifurcation in the phase
portrait at $\mu\ =\ \mu_{r}(g_{1}, odd p)$. Furthermore it is a
saddle-center bifurcation. Again, at this bifurcation point, it is easy
to show that we have spatially
localized but time periodic solutions. These localized solutions can be
pinned at one of the permissible values of $x_{s}$, namely at 
$x_{s}\ = \pm \frac{1} {2},\ \pm \frac{3} {2},\cdots$. These are by
definition discrete breathers\cite{54}. To see it further, we examine
Eq.(4.9). Since at $\mu\ =\ \mu(g_{1}, odd p)$,\  B$_{1}\ =\ 0$, from the
first equation of Eq.(4.9) we find that $\dot{z_{1}}\ =\ 0 $. The
consistent solution is then $z_{1}\ =\ 0$. This in turn
gives $x\ =x_{s}(odd p)\ = \ (l_{1}\ +\ \frac{1} {2})$ where $l_{1}\ =\
0,\ \pm 1,\ \pm2\ \cdots $. Then the second equation of Eq.(4.9) gives
$\dot{z_{2}}\ =\ 0$. This implies that $k\ =\ k_{s}\ =\ (2 l_{2}\ +\
1)\ \pi$ where $l_{2}\ =\ 0, \pm 1, \ \pm2, \ \cdots$. For this
particular case , $k_{s}\ =\ \pi$ or $l_{2}\ =\ 0$. From the perturbative
calculation of 
$\alpha$ appearing in Eq.(3.1)\cite{22}, it can be easily shown that
$\alpha$ is 
proportional to $\tau\ =\ J t$. The energy of the breather in terms
original variables\cite{31} is
\begin{widetext} 
\begin{eqnarray}
\tilde{E}[\mu_{r}(g_{1}, odd p)]\ &=&\ 1 + 2 g_{1}\
\frac{\sinh{\mu_{r}(g_{1}, odd p)}}
{\mu_{r}(g_{1}, odd p)}\  [1\ -\ \mu_{r}(g_{1}, odd p)\
\coth{\mu_{r}(g_{1}, odd
p)}\
+\ \sum_{s\ =\ 1}^{\infty} (-1)^{s}\ \frac{\frac{\pi^{2} s}
{\mu_{r}(g_{1},
odd p)}} {\sinh{\frac{\pi^{2} s} {\mu_{r}(g_{1}, odd p)}}}]\nonumber\\
E_{breather}\ &=&\ H[\mu_{r}(g_{1}, odd p)]\ =\ \frac{4 J} {\lambda}
\sinh{\mu_{r}(g_{1}, odd p)}\ \tilde{E}[\mu_{r}(g_{1}, odd p)].
\end{eqnarray}
\end{widetext}
We also note that for $\frac{J} {\lambda}\ >\ 0 $, this is the
minimum energy spatially localized state. Inasmuch as Eq.(2.5) has a
reflection symmetry, another discrete breather will be obtained by the
transformation, $g_{1}\ \rightarrow\ - g_{1}$. This is discussed later. We
again note that our linearization analysis suggests that at $\mu\ =\
\mu(g_{1}, even p)$, we have another bifurcation. This time it is a  
center saddle bifurcation. However, we find that   
in the region III the phase portraits around $x_{s}\ =\ 0,\ \pm 1, \ \pm
2\ \cdots $ are not that of saddle points. 
In stead these phase diagrams
consisting of closed orbits, are made up of both stable and unstable
manifolds\cite{52}.
So, here we have a relaxation oscillations dynamics as seen in Van der Pol 
oscillators\cite{51, 55}, in the Zeeman models of heartbeat and nerve
impulse\cite{56}. We further note that in
the present problem any relaxation oscillations dynamics is not physically
allowed.  So, these parts of the phase space will remain unaccessible to
the system.

\section{Stationary localized states}
\begin{widetext}
To study stationary localized states of Eq.(2.5), we seek for an
oscillatory
solution in the form\cite{31} 
\begin{equation}
\Psi_{m}(\tau)\ = \ Q_{m}(\tau)\ \exp[i(km-\omega \tau \ +\
\sigma_{0})]\ \exp[- 2 i \tau]  \ , \ \ \ \ \ m \in Z
\end{equation}
where $Q_{m}$ is real and $\sigma_{0}$ is a constant phase factor. From
the real and imaginary parts of Eq.(2.5), we have
\begin{eqnarray}
{(\hat{\Omega} \hat{Q})}_{m}\ &=&\ \omega\  Q_{m}\ + \ \cos{k}\  (1 +\
Q_{m}^{2})\ (Q_{m + 1} + Q_{m - 1})\nonumber\\
&+&\  2\ (1 +\ Q_{m}^{2})\ Q_{m}\ \sum_{j\ =\ 1}^{l} g_{j}
\cos^{2}{k j} \ (Q_{m + j}^{2}\ + \ Q_{m - j}^{2})\ =\ 0;\\
\dot{Q}_{m}\ &+&\ \sin{k}\ (1\ +\ Q_{m}^{2})\  (Q_{m + 1}\ -\ Q_{m -
1})\nonumber\\
&=&\ - 2\ (1 +\ Q_{m}^{2})\ Q_{m}\ \sum_{j\ =\ 1}^{l} g_{j}
\cos{k j} \ \sin{k j}\ (Q_{m + j}^{2}\ -\ Q_{m - j}^{2}) 
\end{eqnarray}
\end{widetext}
where $\hat{Q}$ is the column vector $(Q_{1}, Q_{2}, \cdots\ Q_{m}\
\cdots)$ and $\hat{\Omega}$ is the matrix defined by the left hand side of
Eq.(5.2). Eq.(5.2) with vanishing boundary condition constitutes a
nonlinear
eigenvalue problem for localized states\cite{10, 11,
31}. Eq.(5.3) determines the time
evolution of the localized states. It is a trite algebra to show that
when $g_{j}\ =\ 0$ for all $j$, Q$_{m}(\tau)$ given by Eq.(3.1) is a
solution of Eq.(5.2) and Eq.(5.3). When $k\ =\ 0\ \ {\rm {or}}\ \ \pi,\ \
Q_{m}, \ m\
\in\ Z $ is stationary. A localized state is called "staggered" if $k\ =\
\pi$, and "unstaggered" if $k\ =\ 0$. We
further observe that under the transformation, $Q_{m}\ \rightarrow\
(-1)^{m} Q_{m}$, Eq.(5.2) will remain invariant if  $\omega\ \rightarrow\
- \omega\ \ {\rm{and}}\ \ g_{j}\ \rightarrow\ - g_{j}\ j\ \in l$. This
is indeed in accordance with the reflection symmetry of Eq.(2.5). This
result in turn implies that if an unstaggered localized state, $Q_{m}\
\exp[- i \omega \tau]$ is a solution of Eq.(5.2), the corresponding
staggered localized state, $(-1)^{m} Q_{m}\ \exp[i \omega \tau]$ is then a
solution of the same, provided $g_{j},\ j\ \in\ l$ changes sign for all j.   
To determine locations of these localized states,  we shall specialize on
$l\ =\ 1$ case. This simplification will not affect the merit of the
discussion. From Eq.(5.2), we have
\begin{widetext}
\begin{equation}
\omega\ \  =\ \ - 2 \cos{k}\ -\ (\cos{k}\ +\ 2 g_{1}\ \cos^{2}{k})\
\ \ \frac{\sum_{m} Q_{m}^{2} (Q_{m + 1}\
+\ Q_{m - 1})} {\sum_{m} Q_{m}}\
\ \  -\ \  2 g_{1}\ \cos^{2}{k}\ \frac{\sum_{m} Q_{m}^{3}\ (Q^{2}_{m + 1}\ 
+\ Q^{2}_{m - 1})} {\sum_{m} Q_{m}}. 
\end{equation}  
\end{widetext}
We suppose that $Q_{m}\ >\ 0,\ \ m\ \in\ Z$ and $|g_{1}|\ \ll\ 1$. Then,
the staggered state lies above the phonon band, while the unstaggered
state lies below. We further note that when $\lambda\ \rightarrow\ 0$, $
|g_{1}|\ \rightarrow\ \infty.$ In this case, there is no
localized state, staggered or unstaggered, below the phonon band if
$g_{1}\ <\ 0 $, or above the phonon band for $g_{1}\ > 0$.

The next important point is the effective mass of unstaggered and
staggered localized states\cite{21, 31}. For this we consider the
Hamiltonian, H$_{\rm{eff}}$
given by Eq.(4.5). This consideration will allow us to obtain expressions
for the  effective mass for almost unstaggered and almost staggered states
along with the fully unstaggered and fully staggered localized
states\cite{21}. To calculate the effective mass of nearly unstaggered
localized states, we let $k\
\rightarrow\ 0$ in Eq.(4.5). This in turn yields
\begin{widetext}
\begin{equation}
\lim_{k\ \rightarrow\ 0} H_{\rm {eff}}(x, k, \mu)\ \sim \frac{1} {2}\
m^{-1}_{\rm {eff}}(x)\  k^{2}\  - 2\ A_{0}(\mu) - \ 4 g_{1} A_{1}(\mu)\ +\
8
g_{1}\ A_{2}(\mu)\  \cos{2 \pi x}\ +\ O(k^{4}),
\end{equation}  
\end{widetext} 
where we define
\begin{equation}
 m^{-1}_{\rm {eff}}(x)\ =\ 2\  A_{0}(\mu)\ +\ 8\ g_{1}\ A_{1}(\mu)\ -\ 16\ 
g_{1}\ A_{2}(\mu) \cos{2 \pi x} 
\end{equation}
and A$_{0}$, A$_{1}$, and A$_{2}$ are already defined in the text(see
Eq.(4.7)). We note that $m_{\rm{eff}}(x)$ is even function of both $\mu$
and $x$. It is to be noted that because of the proposed spatial dependence
of $m_{\rm{eff}}$, this is a generalized definition of the quantity. It
can be easily seen that for g$_{1}$ positive semidefinite,
$m_{\rm{eff}}(x)$ is
positive definite for $\mu\ \in\ [- \infty,\ \infty]$. So, when the
coupling constant, $g_{1}\ \ge\ 0$, the effective mass of unstaggered and
nearly unstaggered states is positive definite. 

To obtain the effective mass of staggered localized states, we put $k \ =\
\pi\ -\ \theta$ and then let $\theta\ \rightarrow\ 0 $. This procedure
gives
\begin{equation}
 -\  m^{-1}_{\rm {eff}}(x)\ =\ 2\  A_{0}(\mu)\ -\ 8\ g_{1}\ A_{1}(\mu)\ -\
16\ 
g_{1}\ A_{2}(\mu) \cos{2 \pi x}. 
\end{equation}
Consequently, we get from Eq.(5.7) that when $\mu\ \rightarrow\ 0,\ \
m^{-1}_{\rm{eff}}(x)\ \rightarrow\ - 2$. On the other hand, in the limit 
$\mu\ \gg\ 1$, the asymptotic form of Eq.(5.7) is
\begin{widetext}
\begin{equation}
-\ m^{-1}_{\rm {eff}}(x)\ \sim\ \frac{\exp[\mu]} {\mu}\ [1 - 2\  g_{1}\
\exp[\mu]\ -\ 2 g_{1} \frac{\exp[\mu]} {\mu} (1 + 2 \cos{2 \pi x})].
\end{equation}
\end{widetext}
Let us assume that $2\ g_{1}\ =\ \exp[- (1\ -\ \epsilon) \mu],\ \
\epsilon\ > 0$ and $\epsilon\ \mu\ \rightarrow\ 0$. We
then see from Eq.(5.8) that $m_{\rm{eff}}(x)$ changes sign. So, depending
on the value of $g_{1}$, there will be a critical value of $\mu$,
\ $\mu_{\rm{cr}}(g_{1})$, such that for $\mu\ >\ \mu_{\rm{cr}}(g_{1})$,
staggered as well as nearly staggered localized states will have positive
effective mass. For a more precise analysis of the effective mass, we note
that Eq.(5.7) has two critical values of $\mu$, namely $\mu^{\rm {l}}_{\rm
{cr}},\ \ {\rm{and}}\ \ \mu^{\rm {u}}_{\rm {cr}}$, 
depending on whether $\cos{2 \pi x} \ =\ 1\ \ {\rm{or}}\ - 1$. Again,
these two critical values of $\mu$ move to each other as
$g_{1}$ increases. This result is indeed substantiated by the
numerical analysis of Eq.(5.7) as shown in Fig.5.
It is again to note that for $\mu\ >\ \mu^{\rm{u}}_{\rm{cr}}$,
the effective mass is positive definite, while for $\mu\ <\
\mu^{\rm{l}}_{\rm{cr}}$, the effective mass is negative definite. This in
turn implies that the system will have a strongly localized staggered
state for $|\mu|\ >\ |\mu^{\rm{l}}_{\rm{cr}}(g_{1})|$. We further note
from
Eq.(3.1) that $\mu^{-1}$ gives the measure of
the localization length. So, states for which $\mu\ \rightarrow\ 0 $ in
Eq.(3.1) have long localization lengths. We conclude from this analysis
then that stationary staggered localized states in the vicinity of upper
phonon band edge will have negative effective mass.  
\begin{figure}
\parindent 0.35in
\includegraphics{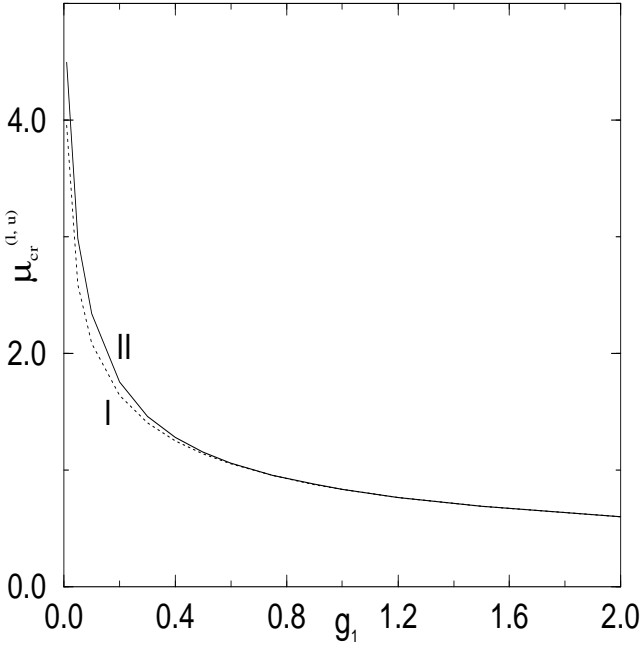}
\caption{ This shows the variation of $\mu^{\rm {l}}_{\rm{cr}}\
{\rm{and}}\ 
\mu^{\rm {u}}_{\rm{cr}}$ as a function of $g_{1}$. Curves I and II
respectively. These critical points
determine
the signature of the effective mass of staggered localized states. Eq(5.7)
in the text.}
\end{figure}

\section{A numerical study of Eq.(2.5)}
\subsection{A study of soliary wave like solutions}
 For the purpose of numerical integration, we first replace
$\Psi_{m},\ m\ =\ 1, 2, 3, \cdots $ by $\Psi_{m} \exp(-2 i \tau)$ in
Eq.(2.5). Next we write $\Psi_{m}\ =\ \Psi_{1m}\ +\ \Im\  \Psi_{2m}$. From
Eq.(2.5), we then get
\begin{widetext}
\begin{eqnarray}
- {\dot\Psi}_{1m}\ &=&\ (1 + \Psi_{1m}^{2}\ +\
\Psi_{2m}^{2}) [(\Psi_{2(m+1)}\
+\ \Psi_{2(m-1)}) + \ 2\ \sum_{j=1}^{l}g_{j}\ \{(\Psi_{1m}
\Psi_{1(m+j)}\nonumber\\  
&+&\ \Psi_{2m} \Psi_{2(m+j)}) \Psi_{2(m+j)} 
+ \ (\Psi_{1m} \Psi_{1(m-j)}\ +\ \Psi_{2m} 
\Psi_{2(m-j)}) \Psi_{2(m-j)}\}]\\ 
{\dot\Psi}_{2m}\ &=&\ (1 + \Psi_{1m}^{2}\ +\
\Psi_{2m}^{2}) [(\Psi_{1(m+1)}\
+\ \Psi_{1(m-1)}) + \ 2\ \sum_{j=1}^{l}\ g_{j}\ \{(\Psi_{1m}
\Psi_{1(m+j)}\nonumber\\  
&+&\ \Psi_{2m} \Psi_{2(m+j)}) \Psi_{1(m+j)} 
 +\ (\Psi_{1m} \Psi_{1(m-j)}\ +\ \Psi_{2m} 
\Psi_{2(m-j)}) \Psi_{1(m-j)}\}].
\end{eqnarray}
\end{widetext}
For the initial condition, we use\cite{8}
\begin{eqnarray}
\Psi_{1m}(\tau\ =\ 0)\ =\ \frac{\sinh {\mu}} {\cosh{m \mu}} \cos{k m}\\ 
\Psi_{2m}(\tau\ =\ 0)\ =\ \frac{\sinh {\mu}} {\cosh{m \mu}} \sin{k m}. 
\end{eqnarray}
The Fourth order Runge-Kutta method is used to integrate these
equations. The chosen time interval for all
calculations is $10^{-4}$. Furthermore, Eq.(2.4) with $\lambda\ =\ 1 $ is
used to check the accuracy of the integration. In this study an one
dimensional lattice,
comprising of 257 lattice points with unit lattice spacing is used. The
initial condition is centered about the middle of the lattice, that is
about the 129th site. For the numerical analysis, we consider two cases,
namely $(i)\ l\ =\ 1$ and $(ii)\ l\ =\ 2$ but $g_{1}\ =\ 0$ in Eq.(2.5). 
For all cases that are presented in this section $\mu\ =\  1$ and $g_{j}\
=\ 0.5,\  j\ =\ 1,\ 2$.

We note that in the first case the permissible values of $k$ are $\pm
\frac{\pi} {2}$ while for the second these values are $\pm \frac{\pi}
{4}\ {\rm{and}}\ \pm \frac{3 \pi} {4}$. 
For the first case the absolute amplitude of the solitary wave like
solution
($Q_{m}$, see Eq.(3.1)) as a function of
site and time (m,\ $\tau$) is shown in Fig.6 for  $k\ =\ \frac{\pi}
{2}$. 
\begin{figure}
\parindent -0.5in
\includegraphics{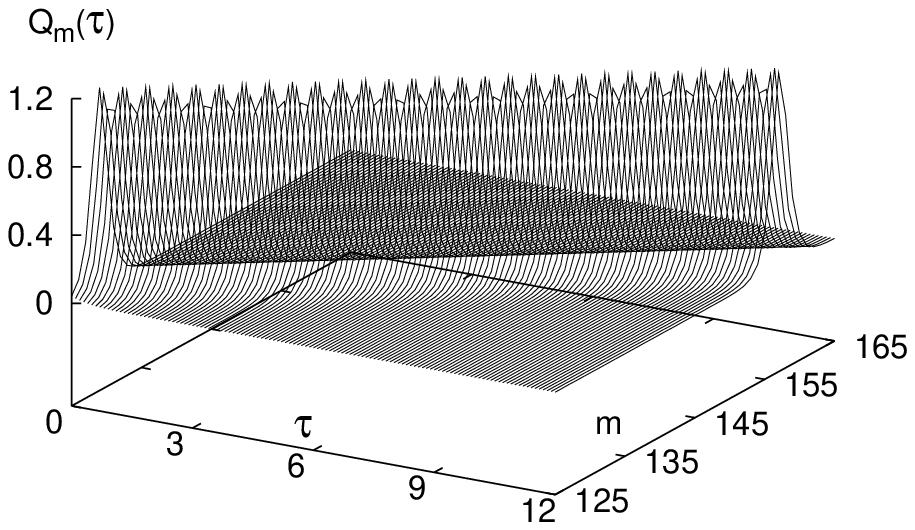}
\caption{The N-AL dynamics of an initial Ablowitz-Ladik soliton
with $k\
=\ \frac{\pi} {2}$, Eq.(2.5). Furthermore, $l\ =\ 1,\ \ g_{1}\ =\ 0.5\
{\rm{and}}\ \mu\ =\ 1.0$. The number of sites in the chain is 257 and
the origin is taken at the center of the chain.}
\parindent 0.3in
\includegraphics{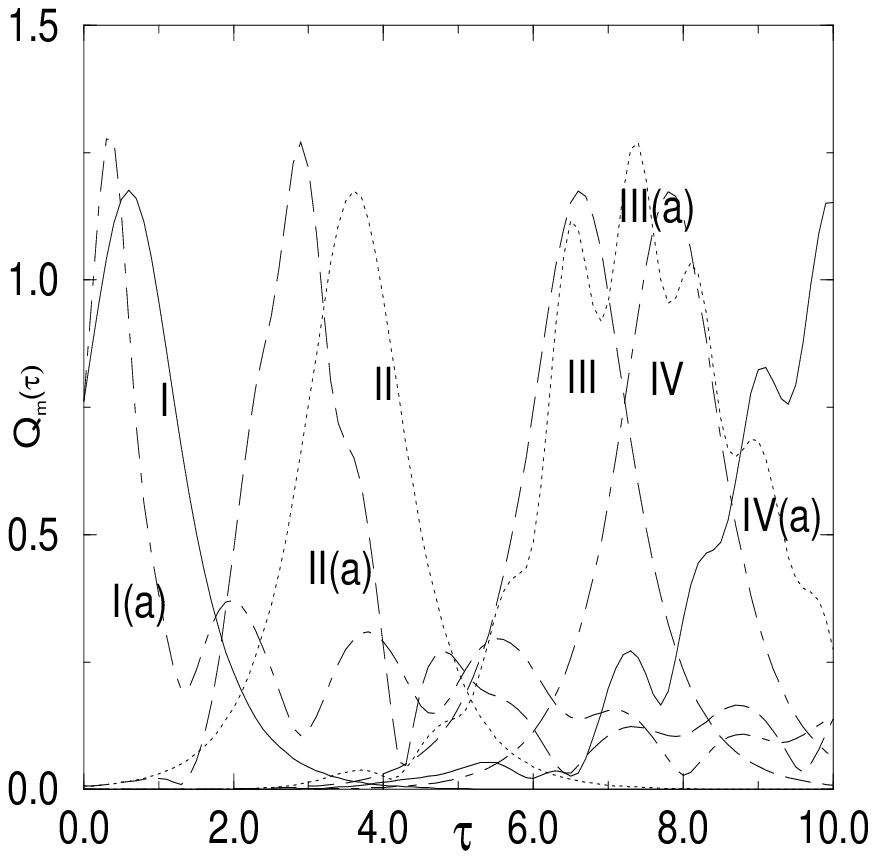}
\noindent
\caption{The N-AL equation (Eq.(2.5)) determined time evolution
of
$Q_{m}(\tau)$ (Eq.(3.1)) for some chosen
values of sites, $m$ .
$k\ =\ \frac{\pi} {2},\ {\rm{and}}\ \frac{\pi} {3}$ respectively. For both
cases $l\ =\ 1,\ \ g_{1}\ =\ 0.5\
{\rm{and}}\ \mu\ =\ 1.0$. The number of sites in the chain is 257 and
the origin is taken at the center of the chain. I, I(a): $m$ = 130,
II, II(a): $m$ = 140, III, III(a): $m$ = 150. The (a) series is for $k\ =\
\frac{\pi} {3}$ while the other one is for $k\ =\ \frac{\pi} {2}$.}
\end{figure}
\begin{figure}
\parindent -0.5in
\includegraphics{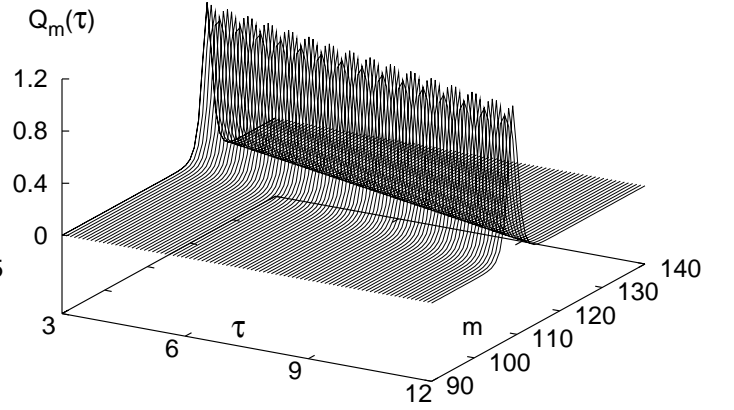}
\caption{The N-AL dynamics of an initial Ablowitz-Ladik soliton
with $k\
=\ - \frac{3 \pi} {4}$, Eq.(2.5). Furthermore, $l\ =\ 2,\
\ g_{1}\ =\ 0.0,\ \ g_{2}\ =\ 0.5\ {\rm{and}}\ \mu\ =\ 1.0$. The number of
sites in the chain is 257 and the origin is taken at the center of the
chain.}
\end{figure}
Furthermore, for this case the time evolution of the absolute
amplitude of the solitary wave ($Q_{m}(\tau)$) for some arbitarily chosen
sites, $m$ are shown in curves I, II, III in Fig.7.
For the second case
$k\ =\ -\ \frac{3 \pi} {4}$ is chosen and the space-time evolution of  the
$Q_{m}(\tau)$ is shown in Fig.8. 
We note that in all cases the initial
profile moves
undistorted. These pictures are expectedly identical
to the space-time evolution of the pure A-L solitons. This is also
verified by numerical integration.  

\subsection{A study of the PN pinning of solitary waves}
Another important aspect that we  study numerically here is the effect of
PN potential, which
arises due to the nonintegrability term in Eq.(2.5) on the space-time
evolution of the initial profile given by Eq.(6.3) and
Eq.(6.4)\cite{31, 41}. We note
that the nonintegrability effect is maximum when $k\ =\ 0$ in Eq.(3.1). On
the other hand it totally vanishes at $|k|\ =\ \frac{\pi} {2}$ for $ l\ =1
$. We study again the propagation of an initial profile for $l\ =\ 1 $, 
but with $\mu\ =\ 1.0$ and $g_{1}\ =\ 0.5$. The accuracy of
the integration is checked by the constancy of
$\mathcal{N}$(Eq.(2.4)). It is found that the loss of constancy becomes
more and
more discernible as $||k|\ -\frac{\pi} {2}|\ \rightarrow \frac{\pi} {2}$.
This makes the simulation for small values of $k$ less reliable.
Fig.9 shows the space-time evolution of $Q_{m}(\tau)$ (Eq.(3.1)) for $k\
=\ \frac{\pi} {3}$.
\begin{figure}
\parindent -0.5in
\includegraphics{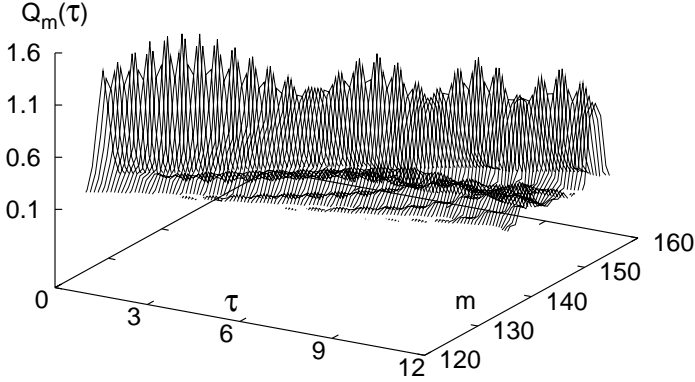}
\caption{The N-AL dynamics of an initial Ablowitz-Ladik soliton
with $k\
=\ \frac{\pi} {3}$, Eq.(2.5). Furthermore, $l\ =\ 1,\ \ g_{1}\ =\ 0.5\
{\rm{and}}\ \mu\ =\ 1.0$. The number of sites in the chain is 257 and
the origin is taken at the center of the chain.}
\end{figure}
\noindent
We note that the dispersion and the distortion of
the initial profile is very transparent in the figure (Fig.9). For better
understanding, we also show through curves I(a) to III(a) in Fig.7 and
I(a) to IV(a) in Fig.10 the time
evolution of the $Q_{m}(\tau)$ for some arbitrarily chosen site or
$m$ values for $k\ =\ \frac{\pi} {3}\ {\rm{and}}\ \frac{\pi} {4}$
respectively. 
\begin{figure}
\parindent 0.25in
\includegraphics{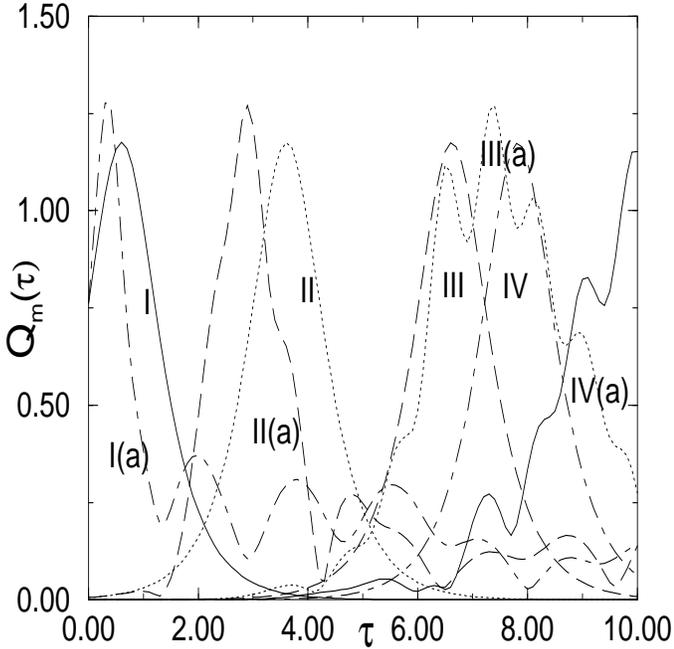}
\caption{Same as Figure 7, but with $k\ =\ \frac{\pi} {4}$ for
both
cases. I, I(a): $m$ = 130, II, II(a): $m$ = 135, III, III(a): $m$ = 140,
IV, IV(a): $m$ = 142. While (a) series is for the N-AL equation, the other
one is for the Ablowitz-Ladik soliton.}
\end{figure}
\noindent 
We note that the distortion of the initial profile and the
intensity of the phonon tail increase as $k$ decreases from $\frac{\pi}
{3}$ to $\frac{\pi} {4}$. This is, of course in the expected
direction. Furthermore, in both cases the initial profile slows down by
leaving phonon behind in the propagation. Through curves I to IV in Fig.10
the time evolution of
$Q_{m}(\tau)$ for same values of $m$ from the Ablowitz-Ladik equation are
included for comparison. This slowing down is very transparent in the
figure (Fig.10). So, it is reasonable to assume that these moving 
profiles will ultimately be trapped. However, we do not observe
the total trapping in our simulations. We possibly need large
integrability breaking term to see the total trapping in the time of
simulations. In case of $k\ =\ \frac{\pi} {2}$, the
nonintegrability term in Eq.(2.5) vanishes, and expectedly the initial
profile (Eq.(6.3) and Eq.(6.4)) propagates undistorted without any change
in its speed and leaving no phonon tail behind as can be seen in Fig.6 and
curves I to III in Fig.7.\\
  
\subsection{A study of interaction of two solitary waves}
An important and also extensively studied field in nonlinear dynamics is
the investigation and the understanding of interaction of two or more
solitary waves of nonintegrable nonlinear equations and the components 
of vector solitons in integrable nonlinear equations. So far only
continuous nonlinear nonintegrable equations are studied\cite{30, 33, 34,
35,36, 37, 38, 43, 44}. This is due to the
nonavailability of any discrete nonlinear equation having solitary wave
like solutions. In this context, therefore, the N-AL equation (Eq.(2.5))
that is proposed here assumes a very great significance. This equation
gives us the opportunity for the first time to study the interaction of
solitary waves of a discrete nonlinear equation. To investigate this
problem, as the initial condition we take the function\cite{35, 36}
\begin{widetext}
\begin{equation}
\Psi_{m}(0)\ =\ \frac{f_{1} \sinh{\mu}} {\cosh[\mu (m - x_{0})]}
\exp[ik_{1}(m - x_{0}]\ +\ \frac{f_{2} \sinh{\mu}} {\cosh[\mu (m +
x_{0})]} \exp[ik_{2}(m + x_{0}] .
\end{equation}
\end{widetext}
We note that the function has two peaks at $x = x_{0}$ and $x = - x_{0}$
respectively. The velocity of the peak at $x_{0}$ is $\frac{\sinh{\mu}}
{\mu} \sin{k_{1}}$ while the velocity of the other peak at $x = - x_{0}$
is $\frac{\sinh{\mu}} {\mu} \sin{k_{2}}$. We further note that when
$|x_{0}|\ \rightarrow\ \infty$, Eq.(6.5) gives two A-L type solitons
(Eq.(3.1)). This analysis is also done
numerically using the fourth order Runge-Kutta method, and the origin is 
placed at the center of the one dimensional chain. For all studies,
we take $ 2 |x_{0}|\ =\ 30$ units, and  $f_{1}\ =\ 
f_{2}\ =\ 1.0$  

The first one is as usual the $l\ =\ 1$ case. For this analysis we take
for all cases $\mu\ =\ 1.0$  $g_{1}\ =\ 0.5$.
Inasmuch as
$ |k|\ =\ \frac{\pi} {2}$ in Eq.(3.1) is an exact one
soliton solution of Eq.(2.5) ,
we consider first the case $k_{2}\ =\ - k_{1}\ =\ \frac{\pi} {2}$. In this
case, we find that two peaks emerge after the collision without any change
in shape and without any emission of phonon (see also ref.34). Of course,
there can be phase shifts in these peaks after collision, but this is not
investigated. This case of collision of solitons is shown in Fig.11. 
\begin{figure}
\parindent -0.5in
\includegraphics{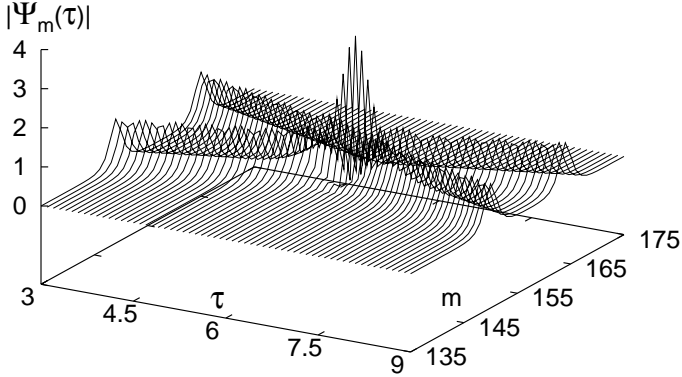}
\caption{The N-AL collision dynamics of two initial A-L
pulses. $k_{2}\
=\ - k_{1}\ =\ \frac{\pi} {2}$. $l\ =\ 1,\ g_{1}\ =\ 0.5\ {\rm{and}}\ \mu\
=\ 1.0$ as mentioned in the text. The number of sites in the chain is 313
and the origin is taken at the middle of the chain.}
\end{figure}
\noindent
We further note that the
result is found to be independent of the value of $g_{1}$. But, we
did not check the effect of changing $\mu$ on the collision. The
other case that is considered is $k_{1}\ =\ -\frac{\pi} {2.05} $ and
$k_{2}\ =\ \frac{\pi} {2}$. When $g_{1}\ =\ 0$, we find that two solitons
collide and after the collision again two peaks emerge without any change
in shape and without any emission of phonon. This is not shown here. On
the other hand, when we give a non-zero value of $g_{1}$ (in our case
0.5), we find the fusion of two solitons on collision. Inasmuch as $k_{2}\
=\ \frac{\pi} {2}$, the direction of the maximum velocity is in the
direction of increasing lattice sites. The fused solution expectedly moves
in that direction. But, at the same time it emits phonons, as shown in
Fig.12\cite{33}. 
\begin{figure}
\parindent -0.5in
\includegraphics{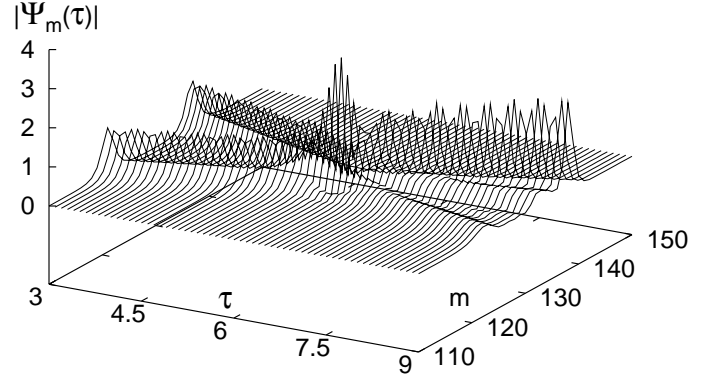}
\caption{Same as in Fig.11, but with $k_{1}\ =\-  \frac{\pi}
{2.05}$,
and\ $k_{2}\ =\ - \frac{\pi} {2}.\ l\ =\ 1,\ g_{1}\ =\ 0.5,\ {\rm{and}}\
 \mu\ =\ 1.0$. In this simulation, the number of sites is 257.}
\end{figure}

The second case that we study is the case of $l\ =\ 2$, but $g_{1}\ =\
0.0$. In this case we have four permissible values of $k$, namely $k\ =\
\pm \frac{3 \pi} {4},\ {\rm{and}}\ \pm \frac{\pi} {4}$. Of course, there
are only two velocities, just like the previous one. However, if choose
$k_{1}\ {\rm{and}}\ k_{2}$ from these allowed values of $k$, for $|x_{0}|\
\rightarrow\ \infty$, Eq.(6.5) will be the solution of Eq.(2.5). In this
analysis, we choose $\mu\ =\ 1.5,\ g_{2}\ =\ 0.5,\ {\rm{and}}\ 2|x_{0}|\ =\ 
30\ {\rm{units}}$. For this case the space-time evolution of a solitary
wave with $k\ =\ - \frac{3 \pi} {4}$ is already shown in Fig.8. In the
first case we take $k_{1}\ =\ -
k_{2}\ =\ -\ \frac{3 \pi} {4}$ and the numerical simulation is shown in
Fig.13. 
\begin{figure}
\parindent -0.5in
\includegraphics{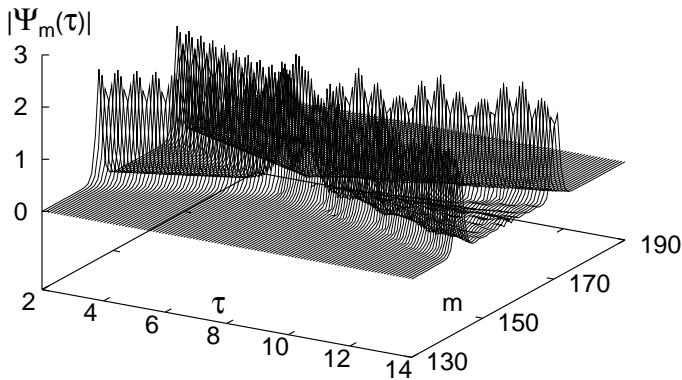}
\caption{The N-AL collision dynamics of two initial A-L
pulses. $k_{2}\ 
=\ - k_{1}\ =\ \frac{3 \pi} {4}$. $l\ =\ 2,\ g_{1}\ =\ 0.0,\ g_{2}\ =\   
0.5\ {\rm{and}}\ \mu\
=\ 1.5$ as mentioned in the text. The number of sites in the chain is 313
and the origin is taken at the middle of the chain.}
\end{figure}
\noindent
In this cae we see that two solitary waves do not pass each
other. On the other hand after coming close to each other, they are
repelled. Other cases that are studied are $k_{1}\ =\ -\
\frac{\pi} {4},\ {\rm{but}}\ k_{2}\ =\ \frac{3 \pi} {4},\ {\rm{and}} 
k_{1}\ =\ -\ \frac{3 \pi} {4},\ {\rm{but}}\ k_{2}\ =\ \frac{ \pi} {4}$
respectively. Our numerical simulations for these cases are shown in 
Fig.14 and Fig.15 respectively. 
\begin{figure}
\parindent -0.5in
\includegraphics{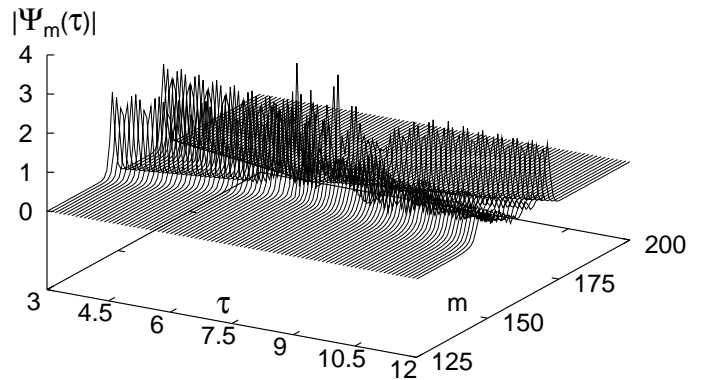}
\caption{ Same as in Fig.13, but with $k_{1}\ =\ - \frac{ \pi}
{4}$,
and\
$k_{2}\ =\ \frac{3 \pi} {4}.\ l\ =\ 2,\  g_{1}\ =\ 0.0,\ g_{2}\ =\ 0.5,\
{\rm{and}}\
 \mu\ =\ 1.5$. In this simulation, the number of sites is 313.}
\end{figure}
\noindent
In the first case, we see that two 
solitary waves fuse after collision and then move in the direction of
positive velocity simultaneously emiting
phonons. In the second case, however, we see the total destruction of the
initial profile, Eq.(6.5) after the collision. 
\begin{figure}
\parindent -0.5in
\includegraphics{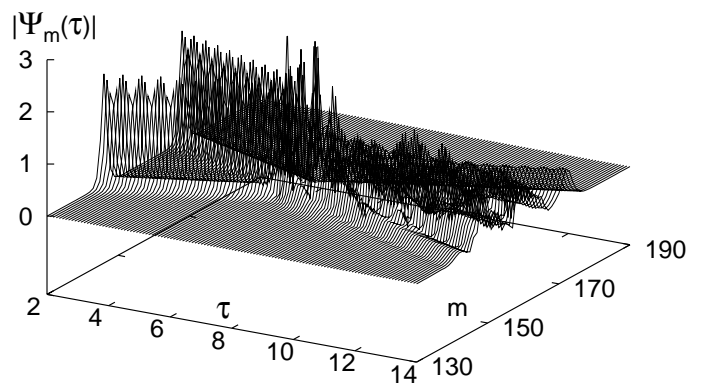}
\caption{Same as in Fig.14, but with $k_{1}\ =\ - \frac{3 \pi}
{4}$,
and\
$k_{2}\ =\ \frac{\pi} {4}.\ l\ =\ 2,\  g_{1}\ =\ 0.0,\ g_{2}\ =\ 0.5,\
{\rm{and}}\ \mu\ =\ 1.5$. In this simulation, the number of sites is 313.}
\end{figure}
This implies that there is
no definite pattern in the collision process. Furthermore, the result of
collision appears to be very sensitive to the phases of the
colliding solitary waves. So, there should be attempts to
understand this collision physics analytically, using perturbative method
and collective coordinate method\cite{37} 

\section{Summary}

A new class of nonintegrable Hamiltonians with tunable nonlinearities is
propsed here to derive a class of (1 + 1) dimensional nonintegrable
discrete nonlinear Schr$\ddot{\rm{o}}$dinger equations, which include both
some known nonlinear equations such as Ablowitz-Ladik nonlinear
Schr$\ddot{\rm{o}}$dinger equation\cite{6, 7}, IN-DNLS\cite{31, 45, 46} 
and a new subset, collectively
christened as the N-AL equation(Eq.(2.5)). The relevant equations are
obtained from the proposed Hamiltonian by the standard procedure of
classical mechanics, but by employing a standard generalized definition of
Poisson brackets, as shown
in the Appendix A below. In this paper two cases of the N-AL equation are
investigated analytically as well as numerically. These cases are
$(i)\ l\ = 1,\ {\rm{and}}\ (ii)\ l\ =\ 2,\ {\rm{but}}\ g_{1}\ =\ 0.0$
in Eq.(2.5). 

An important characteristic of the N-AL equation is shown to be that it
can allow spatially localized states of A-L type (Eq.(3.1)) for certain
permissible values of the parameter $k$ to travel without distortion and
without emiting phonons. This is found both analytically as well as 
numerically. So, these special solutions have in them the required
balance of the nonlinearity and dispersion\cite{5} and also are
transparent to the PN potential arising from the lattice
discreteness\cite{31, 41}. The transparency
of these solutions to the PN potential proves that these have the
{\it{continuous}} translational symmetry
of the one-soliton solution of the Ablowitz-Ladik equation
(Eq.(3.1))\cite{31}. 

Trapped and moving solitons are found from Eq.(2.5). This analysis is done
using a standard perturbative procedure\cite{21, 22}. The most interesting
part of this
analysis, however, is that it shows the existence of {\it{saddle center
bifurcations}}. This in turn implies the presence of
minimum energy breathers in the system. Breathers are by definition time
periodic,
spatially localized solutions of equations of motion for classical
degrees of freedom interacting on a lattice\cite{54}. We note that
eq.(4.1) satisfy these requirements.
Since, the saddle center bifurcation is seen in the perturbative method,
it is, therefore, imperative to study the existence of breathers in the
system numerically. We further note that in this system we also find a
chain of centers. The origin of this situation is explained in the text. 
One interesting consequence of this occurence is phase portraits showing
relaxation oscillations in regions, which are physically forbidden
parts of the phase space for this system. Since phase portraits are very
rich in characteristics, it is worthwhile to study them quite elaborately
by varying extensively parameters, such as $\mu\  {\rm{and}}\ g_{1}$ and 
and also including more terms in $S(\mu, x)$ (Eq.(4.1). 

The presence of stable localized states in Eq.(2.5) is also investigated, 
albeit a more systematic analysis, presumably by the discrete variational
method is required. This work is indeed in progress. The effective mass of
localized states are analyzed here using the effective Hamiltonian, given
by Eq.(4.5)\cite{21}. It is found that nearly unstaggered as well as
unstaggered localized states will always have positive effective mass. On
the other hand, nearly staggered and staggered localized states can have
both negative and positive effective depending on the value of $\mu$ in
Eq.(3.1). However, large width but small amplitude nearly staggered and
staggered localized states do have the expected negative effective mass.

It is further shown here numerically that all other solitary wave profiles
(Eq.(6.3) and Eq.(6.4)) under the dynamics described by Eq.(2.5) and 
for which the nonintegrable term in Eq.(2.5) does not vanish, suffer
distortion and emit phonons in the propagation. Further investigations
along this line by altering the strength of the nonintegrable term by
altering the value of $k$ reveal that the extent of distortion and the
extent of emission of phonons by the initial solitary wave profiles
(Eq.(6.3) and Eq.(6.4)) depend on the magnitude of the nonintegrable 
term. We conclude from these two findings that any general solitary wave
profile due to the presence of the nonintegrable term in Eq.(2.5) is
subjected to both the dynamical imbalance between the nonlinearity and the
dispersion, and to the PN potential. We note that the 
continuous translational symmetry of these solitary wave 
profiles breaks down due to the nonintegrable term in
Eq.(2.5). Consequently, these profiles scan the lattice discreteness in
the propagation. This, in turn implies that the moving profiles experience
the effect of PN potential in the dynamics. Since, any general solitary
wave profile suffers from these two factors, we expect that the profile
can either be trapped by emiting
phonons or the localized structure will spread uniformly due to the effect
of extra dispersion. Notwithstanding the slowing down of initial profiles
are observed in our simulations, our numerical study in this matter,
however, is not conclusive. So, a more elaborate study increasing the
coupling constants in the nonintegrable term in Eq.(2.5) and also changing
the parameter $\mu$ in Eq.(3.1) is needed. This will be done in future.   

Finally, the N-AL equation (Eq.(2.5))is the discrete analog of the famous
$\phi^{4}$
equation, however with an important difference, that this equation only
gives dynamical solitary wave like solutions. None the less, the
importance of the N-AL equation in this regard cannot be overlooked. This
equation gives the opportunity for the first time to study the collision
dynamics of solitary wave profiles using a discrete nonlinear
nonintegrable equation. Here only the collision of two solitary wave
profiles are studied (Eq.(6.5)). In this study we find that at least the
interaction of two solitary wave profiles have some universal features,
meaning that the outcome does not depend critically on the nature of the
nonlinear equation. It is also equally true that we do not find so far any
systematic pattern in the dynamics. It is, therefore, necessary to study
this problem in details both analytically and numerically to discern if
there is any specific pattern in the dynamics. Another imporatnt aspect to
study is the fractal structure of the emeging profile after the collision
as a function of the initial velocity and the strength of the
nonintegrable term\cite{37, 38}.

We conclude by noting that this study offers a very significant insight
into the transport properties of the DHM. Furthermore, the N-AL equation
can be used to study transport properties of localized states in soft
molecular chains, having the coexsistence of nonlinearity and disorder,
nolinearity and quasiperiodicity or incommesurabily in various parameters,
such as the hopping integral, J and the nonintegrable coupling constants , 
$g_{j},\ j\ =\ 1,\ 2,\ \cdots$. Furthermore, the aplicability of the  N-AL
equation in the  study exciton dynamics in photosynthesis and in molecular
solids should be explored.

{\appendix

\section{Generalized Poisson Bracket}
We consider a dynamical system having 2 N generalized coordinates,
$\{\phi_{n},\ \phi^{\star}_{n}\},\ n\ =\ 1,\ \dots N $. Let U and V be
any two general dynamical variables of the system. We now define a
nonstandard Poisson bracket\cite{2}
\begin{equation}
\{U,\ V \}_{\{\phi,\ \phi^{\star}\}}\ =\ \sum_{n\ =\ 1}^{N}\
(\frac{\partial U} {\partial \phi_{n}}\  \frac{\partial V} {\partial
\phi^{\star}_{n}}\ -\ \frac{\partial V} {\partial \phi_{n}}\
\frac{\partial U} {\partial \phi^{\star}_{n}})\ (1\ +\ \lambda\
|\phi_{n}|^{2}).
\end{equation}
When $U\ =\ \phi_{m}\ {\rm {and}}\  V\ =\ \phi^{\star}_{l}$, we then have
\cite{21, 31, 57}
\begin{equation}
\{\phi_{m},\  \phi^{\star}_{l} \}_{\{\phi,\ \phi^{\star}\}}\ =\
(1\ +\ \lambda\ |\phi_{m}|^{2})\ \delta_{ml}.
\end{equation}
From Eq.(A1) we further have for $\{m,\ l\}$
\begin{equation}
\{\phi_{m},\  \phi_{l} \}_{\{\phi,\ \phi^{\star}\}}\ =\
\{\phi^{\star}_{m},\  \phi^{\star}_{l} \}_{\{\phi,\ \phi^{\star}\}}\ =\ 0.
\end{equation}
Then when $U\ =\ \phi_{m}$ and $V\ =\ {\rm {H}}$, we have from
Eq.(A1)
\begin{equation}
\{\phi_{m}, {\rm {H}} \}_{\{\phi,\ \phi^{\star}\}}\ =\ (1\ +\ \lambda
\ {|\phi_{m}|^{2}})\ \frac{\partial {\rm {H}}} {\partial
{\phi^{\star}_{m}}}.
\end{equation}
We then write for the dynamical evolution of the m-th generalized
coordinate, $\phi_{m}$\cite{31, 57}
\begin{equation}
i\ \frac{d \phi_{m}} {d t\ \ \ }\ =\ \{\phi_{m}, {\rm {H}} \}_{\{\phi,\
\phi^{\star}\}}\ =\ (1\ +\ \lambda
\ {|\phi_{m}|^{2}})\ \frac{\partial {\rm {H}}} {\partial
{\phi^{\star}_{m}}}.
\end{equation}
This is consistent because
\begin{eqnarray}
i\ \frac{d U} {d t\ \ }\ &=&\ \sum_{n\ =\ 1}^{N} (\frac{\partial U}
{\partial
\phi_{n}}\ i\ \dot{\phi_{n}}\ +\ \frac{\partial U} {\partial
\phi^{\star}_{n}}\ i\ \dot{\phi^{\star}_{n}})\nonumber\\
&=&\ \sum_{n\ =\ 1}^{N} (1\ +\ \lambda
\ {|\phi_{n}|^{2}})\ (\frac{\partial {\rm {U}}} {\partial
\phi_{n}}\ \frac{\partial {\rm {H}}} {\partial \phi^{\star}_{n}}
\ -  \ \frac{\partial {\rm {U}}} {\partial \phi^{\star}_{n}}\
\frac {\partial {\rm {H}}} {\partial \phi_{n}})\nonumber\\
&=&\ \{U,\ {\rm {H}}\}_{ \{\phi,\ \phi^{\star}\}}.
\end{eqnarray}
We note now that when $U\ =\ {\rm{H}}$, we have $\frac{d {\rm{H}}} {d t\ 
}\ 
=\ 0$. In other words, ${\rm{H}}$ is a constant of motion. Consider
${\mathcal{N}}$ given by Eq.(2.4). To show that it is a constant of motion,
we write ${\rm{H\ =\ H_{0}\ +\ H_{1}}}\ -\ \frac{2} {\lambda} (\frac{\nu}
{\lambda}\ - J)\ {\mathcal{N}}$, where we define
\begin{eqnarray}
{\rm {H_{0}}}\ &=&\ J\ \sum_{n}\ (\phi^{\star}_{n}\ \phi_{n+1}\ +\  
\phi^{\star}_{n+1}\ \phi_{n})\ + \frac{2 \nu} {\lambda}\ \sum_{n}
|\phi_{n}|^{2};\\
{\rm{H_{1}}}\ &=& \ -\ \frac{1} {2}\ \sum_{n}\ \sum_{j\ =\
1}^{l} g^{j}_{0}\ (\phi^{\star}_{n}\ \phi_{n+j}\ +\
\phi^{\star}_{n+j}\ \phi_{n})^{2}. 
\end{eqnarray}
We then find that
\begin{equation}
i\ \frac{d {\mathcal{N}}} {d t\ \ }\ 
=\ \lambda\ \sum_{n}  
[(\phi^{\star}_{n}\ \frac{\partial {\rm {H_{0}}}} {\partial
\phi^{\star}_{n}}
\ -  \  \phi_{n}\
\frac {\partial {\rm {H_{0}}}} {\partial \phi_{n}})
+\   
(\phi^{\star}_{n}\ \frac{\partial {\rm {H_{1}}}} {\partial
\phi^{\star}_{n}}\ -  \  \phi_{n}\
\frac {\partial {\rm {H_{1}}}} {\partial \phi_{n}})].
\end{equation}
By using (A7) and (A8), it is very simple to show that the right hand
side of (A9) is zero. Hence, ${\mathcal{N}}$ is a constant of motion.
}


\begin{thebibliography}{99}

\bibitem{1} P. G. Drazin and R. S. Johnson, {\it Solitons
: an
introduction} (Cambridge Univ. Press, Cambridge, 1989).

\bibitem{2} A. Scott, {\it Nonlinear Science : Emergence ${\mathcal
{\&}}$
Dynamics of Coherent Structures}, (Oxford University Press,
U. K. 1999).

\bibitem{3} See e. g. M. Remoissenet, {\it Waves Called Solitons
: Concepts and
Experiments} (Springer-Verlag, Berlin, 1996).

\bibitem{4} Yu. S. Kivshar and B. A. Malomed, Rev. Mod. Phys. {\bf{63}},
761, (1989).

\bibitem{5} R. Z. Sagdeev, S. S. Meiseev, A. V. Tur and V. V. Yanevskii, 
{\it Nolinear Phenomena in Plasma Physics and Hydrodynamics},
ed. R. Z. Sagdeev, (Mir Publishers, Moscow), 137 (1986).  

\bibitem{6} M. J. Ablowitz and P. A. Clarkson, {\it
Solitons, Nonlinear
Evolution Equations and Inverse Scattering} (Cambridge
Univ. Press, Cambidge, 1991). 

\bibitem{7} M. J. Ablowitz and J. L. Ladik,
J. Math. Phys. {\bf 16}, 598,  (1975), {\it {ibid}}, {\bf 17}, 1011,
(1976).

\bibitem{8} S. Takeno and S. Homma, J. Phys. Soc. Jpn, {\ 60}, 731
(1991).

\bibitem{9} E. W. Laedke, K. H. Spatschek and S. K. Turitsyn,
Phys. Rev. Lett. {\bf 73}, 1055 (1994). 

\bibitem{10} A. Ghosh, B. C. Gupta and
K. Kundu, J. Phys. : Condens. Matter, {\bf 10}, 2701 (1998).

\bibitem{11} K. Kundu and B. C. Gupta, Eur. Phys. J. B {\bf{3}} 23 (1998);
B. C. Gupta and K. Kundu, {\it Nonlinear Dynamics: Integrabilty and Chaos}
Eds. M. Daniel, K. M. Tamizhmani and R. Sahadevan, p. 193 (Narosa, New
Delhi, 2002)

\bibitem{12} V. E. Zakharov and A. B. Shabat, Zh. Eksp. Teor. Fiz. {\bf
61}, 118 (1971), [Sov. Phys. JETP {\bf 34},
62 (1972)].

\bibitem{13}  A. S. Davydov and N. I. Kislukha, Zh. Eksp. Teor. Fiz. {\bf
71}, 1090 (1976), [Sov. Phys. JETP {\bf 44}, 571 (1976)].

\bibitem{14} A. C. Scott, Phys. Rev. A {\bf 26}, 578 (1982)

\bibitem{15}  A. Scott, {\it Davydov's Soliton}, Phys. Rep. {\bf 217}, 
3 (1992).

\bibitem{16} J. C. Bronski, L. D. Carr, 
B. Deconinck and J. N. Kutz, Phys. Rev. Lett. {\bf 86}, 1402 (2001),\\
A. Trombettoni and A. Smerzi, Phys. Rev. Lett. {\bf 86}, 2353 (2001).

\bibitem{17}  I. V. Barashenkov and E. V. Zemlyanaya,
Phys. Rev. Lett. {\bf 83}, 2568 (1999).

\bibitem {18}   Ch. Claude, Y. S. Kivshar, O. Kluth and
K. H. Spatchek, Phys. Rev. B {\bf 47}, 14228 (1993).

\bibitem{19} A. B. Aceves, C. De Angelis, T Peschel, R. Muschall, F.   
Lederer, S. Trillo and S. Wabnitz, Phys. Rev. E {\bf 53}, 1172 (1996).

\bibitem{20}  S. Yomosa, J. Phys. Soc. Jpn. {\bf 52}, 1866 (1983).

\bibitem{21} K. Kundu, Phys. Rev. E, {\bf 61}, 5839 (2000).

\bibitem{22} A. A. Vakhnenko and Yu. B. Gaididei, Theo. Math. Fiz. {\bf
68}, 350 (1986) [Theor. Math. Phys. {\bf 68}, 873 (1987)].

\bibitem{23} S. Takeno, J. Phys. Soc. Jpn. {\bf 61}, 1433 (1992).

\bibitem{24} K. Hori and S. Takeno, J. Phys. soc. Jpn. {\bf 61} 4263
(1992).

\bibitem{25}V. V. Konotop and S. Takeno, Phys. Rev. B {\bf 55}, 11342
(1997).

\bibitem{26} Yu. S. Kivshar, D. E. Pelinovsky, T. Cretegny and M. Peyard,
Phys. Rev. Lett. {\bf 80}, 5032 (1998).

\bibitem{27} A. S. Davydov, Zh. Eksp. Teor. Fiz. {\bf 78},
789 (1980) [Sov. Phys. JETP {\bf 51}, 397 (1980)].

\bibitem{28} H. Bolterauer, {\it Davydov's soliton revisited, self-
trapping of vibrational energy in protein}, eds. P. L. Christeinsen and
A. C. Scott, NATO ASI series, Vol. {\bf 234}, 309 (1990).

\bibitem{29} W. F{$\ddot{\rm{o}}$}rner and J. Ladik, {\it Davydov's
soliton revisited, self-
trapping of vibrational energy in protein}, eds. P. L. Christeinsen and
A. C. Scott, NATO ASI series, Vol. {\bf 234}, 267 (1990).

\bibitem{30} S. V. Dmitriev, Yu. S. Kivshar and T. Shigenari,
Phys. Rev. Lett., {\bf 64}, 056613 (2001).

\bibitem{31} D. Cai, A. R. Bishop,  N. Gr${\o}$nbech-Jensen,
Phys. Rev. Lett. {\bf 72}, 591 (1994). 

\bibitem{32} M. Johansson and Yu. S. Kivshar, Phys. Rev. Lett. {\bf 82},
85 (1999).

\bibitem{33} R. K. Bullough and P. J. Caudrey, {\it Topics in Current
Physics : Solitons}, Eds. R. K. Bullough and P. J. Caudrey, p. 6
(Springer-Verlag, Berlin, 1980).

\bibitem{34} S. Aubry, J. Chem. Phys. {\bf 64}, 3392 (1976).

\bibitem{35} A. E. Kudryavtsev, Pis'ma Zh. Eksp. Teor. Fiz. {\bf 22},
178 (1975), [JETP Lett., {\bf 22}, 82 (1975)].

\bibitem{36} B. S. Getmanov,  Pis'ma Zh. Eksp. Teor. Fiz. {\bf 24},
323 (1976), [JETP Lett. {\bf 24}, 291 (1976)].

\bibitem{37} P. Anninos, S. Oliveira and R. A. Matzner, Phys. Rev. D {\bf
44} 1147 (1991).

\bibitem{38} J. Yang and Yu Tan, Phys. Rev. Lett. {\bf 85}, 3624 (2000),
Phys. Lett. A {\bf 280}, 129 (2001), Yu Tan and J. Yang, Phys. Rev. E. 
{\bf 64}, 056616 (2001).

\bibitem{39} D. H. Dunlop,  H-L Wu and P. Phillips,
Phys. Rev. Lett. {\bf 65}, 88 (1990).
 
\bibitem{40} K. Kundu, D. Giri and K. Ray,
J. Phys. A : Math Gen. {\bf 29}, 5699 (1996).

\bibitem{41} Y. S. Kivshar and D. K. Campbell, Phys. Rev. E
{\bf 48}, 3077 (1993).

\bibitem{42} S. Takeno, {\it Davydov's soliton revisited, self-
trapping of vibrational energy in protein}, eds. P. L. Christeinsen and
A. C. Scott, NATO ASI series, Vol. {\bf 234}, 31 (1990).

\bibitem{43} G. I. Stegeman and M. Segev, Science, {\bf 286}, 1518 (1999),
M. Segev and G. Stegeman, {\it Physics Today}, 42 (August, 1998).

\bibitem{44} T. Kanna and M. Lakshmanan, Phy. Rev. Lett. {\bf 86}, 5043
(2001), C. Anastassiou, M. Segev, K. Steiglitz, J. A. Giordmaine and
M. Mitchell, Phys. Rev. Lett. {\bf 83}, 2332 (1999).

\bibitem{45} M. Salerno, Phys. Rev. A {\bf 46}, 6856 (1992).

\bibitem{46} V. V. Konotop and M. Salerno, Phys. Rev. E {\bf 55}, 4706
(1997), {\it ibid} {\bf 56}, 3611 (1997).

\bibitem{47} E. I. Rashba, in {\it Excitons} eds. E. I. Rashba and
M. D. Sturge (North Holland Publ, Amsterdam, 1982).

\bibitem{48} Y. Toyozawa, In {\it Molecular Aggregates} {\it Springer
series in Solid- State Sciences}. 49, eds. P. Reineker, H. Haken and
H. C. Wolf
(Springer-Verlag, Berlin, 1983).

\bibitem{49} H. van Amerongen, L. Valkunas and R. van Grondelle,
{\it Photosynthetic EXCITONS}, Chap. 6, pp.197-240, (World Scientific,
Singapore, 2000). 

\bibitem{50} D. I. Kaup, SIAM J. Appl. Math. {\bf 31}, 121 (1976).

\bibitem{51}D. K. Arrowsmith and C. M. Place, {\it Ordinary Differential
Equations}, (Chapman and Hall, London, 1982).

\bibitem{52} K. T. Alligood, T. D. Sauer and J. A. Yorke, {\it Chaos: an
introduction to dynamical systems}, (Springer-Verlag, New York, 1997).

\bibitem{53} P. G. Drazin, {\it Nonlinear Systems}, (Cambridge University
Press U. K. 1992).

\bibitem{54} S. Flach, K. Kladko, and R. MacKay, Phys. Rev. Lett. {\bf
78}, 1207 (1997).

\bibitem{55} B. Meerson and G. I. Shinar, Phys. Rev. E {\bf 56}, 256
(1997).

\bibitem{56} E. C. Zeeman, {\it Differential Equations for the Heartbeat
and Nerve Impulse}, Salvador Symposium on Dynamical Systems, Academic
Press, pp. 683-741. 

\bibitem{57} A. Das, {\it Integrable Models}, (World Scientific,   
Singpore, 1989).

\end{thebibliography}
\end{document}